\documentstyle[preprint,tighten,eqsecnum,aps,floats,psfig,epsfig]{revtex}

\begin{document}
\draft
\title{
The three-dimensional randomly dilute Ising model: \\ Monte Carlo results
}
\author{Pasquale Calabrese,$^1$ Victor Mart\'\i n-Mayor,$^{2,3}$
       Andrea Pelissetto,$^4$ Ettore Vicari$^5$ }
\address{$^1$ Scuola Normale Superiore and  INFN, Piazza dei Cavalieri 7,
 I-56126 Pisa, Italy.}
\address{$^2$ Departamento de F\'\i sica Teorica I,
    Universidad Complutense de Madrid, \\
   E-28040 Madrid, Spain}
\address{$^3$ Instituto de Biocomputaci\'on y F\'{\i}sica de
    Sistemas Complejos (BIFI), Universidad de Zaragoza, E-50009 Zaragoza,
    Spain}
\address{$^4$ Dip. Fisica dell'Universit\`a di Roma ``La Sapienza" \\
and INFN, P.le Moro 2, I-00185 Roma, Italy}
\address{$^5$ Dip. Fisica dell'Universit\`a di Pisa
and INFN, V. Buonarroti 2, I-56127 Pisa, Italy}
\address{
{\bf e-mail: \rm 
{\tt calabres@df.unipi.it},
{\tt Victor@lattice.fis.ucm.es},
{\tt Andrea.Pelissetto@roma1.infn.it},
{\tt vicari@df.unipi.it}
}}

\date{\today}

\maketitle

\begin{abstract}
We perform a high-statistics simulation of the three-dimensional 
randomly dilute Ising model on cubic lattices $L^3$ with $L\le 256$.
We choose a particular value of the density, $x=0.8$,
for which the leading scaling corrections are suppressed.
We determine the critical exponents, obtaining 
$\nu = 0.683(3)$, $\eta = 0.035(2)$,
$\beta = 0.3535(17)$, and $\alpha = -0.049(9)$, in agreement
with previous numerical simulations.
We also estimate numerically the fixed-point values of the 
four-point zero-momentum couplings that are used in field-theoretical 
fixed-dimension studies. Although these results
somewhat differ from those obtained using 
perturbative field theory, the field-theoretical estimates 
of the critical exponents do not change
significantly if the Monte Carlo result for the fixed point is used. 
Finally, we determine the six-point zero-momentum couplings,
relevant for the small-magnetization expansion of the equation of state, and 
the invariant amplitude ratio $R^+_\xi$ that 
expresses the universality of the free-energy density per 
correlation volume. We find $R^+_\xi = 0.2885(15)$.
\end{abstract}




\section{Introduction.}
\label{intro}

During the last few decades many theoretical and experimental
studies have investigated the critical properties
of statistical systems in the presence of quenched disorder.
Typical examples are randomly dilute uniaxial antiferromagnets,
for instance,
Fe$_x$Zn$_{1-x}$F${}_2$ and  Mn$_x$Zn$_{1-x}$F${}_2$,
obtained by mixing a uniaxial antiferromagnet with short-range interactions
with a nonmagnetic material.  Experiments 
find that, for sufficiently low impurity concentration 1--$x$,
these systems undergo a second-order phase transition
at $T_c(x) < T_c(x=1)$, with critical exponents independent of 
the impurity concentration.
The experimental results have been summarized in Ref.~\cite{Belanger-00},
which reports $\alpha=-0.10(2)$, $\nu=0.69(1)$, and $\beta=0.350(9)$.
These estimates are definitely different from the
values of the critical exponents of the pure Ising universality class,
where, e.g., $\alpha=0.1096(5)$ (Ref.~\cite{CPRV-02}), and thus 
indicate that the impurities change 
the nature of the transition that belongs to a new random 
universality class.  In the presence of an external magnetic field, 
dilute uniaxial antiferromagnets show a different critical 
transition, belonging to the universality class of the 
random-field Ising model \cite{FA-79,Cardy-84,Belanger-98,crossexp}.

A simple model for dilute uniaxial systems is provided by 
the three-dimensional random Ising model (RIM) with Hamiltonian
\begin{equation}
{\cal H}_x = J\,\sum_{<ij>}  \rho_i \,\rho_j \; s_i s_j,
\label{latticeH}
\end{equation}
where the sum is extended over all nearest-neighbor sites,
$s_i$ are Ising spin variables, and
$\rho_i$ are uncorrelated quenched random variables, which are equal to one 
with probability $x$ (the spin concentration) and zero with probability $1-x$
(the impurity concentration). 
For small $1-x$, i.e. above the percolation threshold of the spins, 
this model shows a critical transition analogous 
to that observed in experiments and whose nature has been the 
object of many theoretical studies, 
see, e.g., Refs. \cite{Aharony-76,Stinchcombe-83,PV-r,FHY-01}.

Numerical Monte Carlo simulations
\cite{SST-95,WD-98,BFMMPR-98,Hukushima-00,MGI-00,BB-01,BCBJ-02}
had long been inconclusive in setting the question of the critical behavior
of the RIM. While the measured critical exponents were definitely different
from the Ising ones, results apparently depended on the spin concentration,
in disagreement with renormalization-group (RG) theory. Only recently has the 
question been clarified. Ref.~\cite{BFMMPR-98} showed the presence of 
very strong concentration-dependent scaling corrections with 
exponent $\omega = 0.37(6)$. Only if they are properly taken into account, the 
numerical estimates of the critical exponents become dilution independent as
expected.  Their final estimates are
$\nu=0.6837(53)$ and $\eta=0.0374(45)$,
from which one also derives $\beta=0.3546(28)$ and $\alpha=-0.051(16)$
using scaling relations.
These results are in good agreement with the experimental ones reported above, 
although the numerical estimate of $\alpha$ is slightly different.

Randomly dilute Ising systems can also be studied by using 
the field-theoretical (FT) approach \cite{PV-r,FHY-01}.
The starting point is the cubic-symmetric Hamiltonian \cite{GL-76}
\begin{equation}
{\cal H}_{\rm LGW} =  \int d^d x \Bigl\{ {1\over 2} \sum_{i=1}^{N}
      \left[ (\partial_\mu \phi_i)^2 +  r \phi_i^2 \right]  
+{1\over 4!} \sum_{i,j=1}^N \left( u_0 + v_0 \delta_{ij} \right)
\phi^2_i \phi^2_j  \Bigr\}, 
\label{Hphi4rim}
\end{equation}
where $\phi_i$ is an $N$-component field.
By using the standard replica trick,
it can be shown that in the formal limit $N\to 0$ such a model corresponds to 
a system with quenched disorder effectively coupled to the energy density,
as is the case of the RIM \cite{GL-76}. As is well known, the limit $N\to 0$ is
a subtle one. In the standard perturbative approaches, the limit is taken
naively---we simply set $N=0$ in the perturbative expansion---implicitly 
assuming that the replica symmetry is not broken.
In recent years, however, this assumption has been
questioned~\cite{DHSS-95} on the ground that 
the RG approach may not take into account other  
local minimum configurations of the random Hamiltonian (\ref{Hphi4rim}), 
which may cause the spontaneous breaking of the replica symmetry.
However, a fixed-dimension perturbative two-loop calculation \cite{PPF-01} 
in a perturbative approach proposed in Ref.~\cite{DHSS-95} finds that the 
standard replica-symmetric fixed point is stable under any
replica-symmetry breaking perturbation, thereby supporting the standard
approach. In this paper,
we do not further consider this issue and in the following 
we always assume that the standard approach is correct. Note that the good 
agreement among numerical and field-theoretical results supports this 
assumption, although one cannot exclude that replica-symmetry breaking 
effects can only be seen very close to the critical point.

In the FT approach one looks for stable fixed points in the region $u_0<0$ 
(or, equivalently $u < 0$).
If the pure fixed point at $u=0$ is stable, disorder is irrelevant, while 
the presence of a new stable fixed point with $u < 0$, indicates that 
disorder is relevant and that dilute systems belong to a new 
universality class. Numerical and experimental results indicate 
that in dilute Ising systems the correct scenario is the second one and 
thus a new random fixed point should be present with $u^*<0$.

The most precise FT estimates of critical quantities are presently obtained by 
using perturbative methods. However, in the case of random systems the 
perturbative approach faces new difficulties: the perturbative series
are not only divergent, but are also non-Borel 
summable \cite{BMMRY-87,AMR-99}. This means that even the knowledge 
of the complete perturbative series does not allow the exact computation of 
the critical quantities. These difficulties are clearly present in 
the $\sqrt{\epsilon}$-expansion and in the related minimal-subtraction
scheme without $\epsilon$-expansion \cite{SAS-97,FHY-00,BH-03}. 
The expansion in $\sqrt{\epsilon}$ is not well behaved and does not allow 
quantitative determinations of the critical exponents, while in the 
minimal-subtraction scheme results are very sensitive to the 
resummation method. If the Chisholm-Borel method is used \cite{FHY-00},
no random fixed point is found with the 
longest available series (five loops).
Apparently, four-loop series provide the most accurate results and increasing
the length of the expansion does not help improving the precision of the 
results. On the other hand, if a double Pad\'e-Borel resummation is used 
as proposed in Ref.~\cite{AMR-99}, a random fixed point is found also 
at five loops \cite{BH-03}. The estimates of the critical exponents 
are in any case not very precise, and moreover, at variance with the 
fixed-dimension approach described below, the stability-matrix eigenvalues
turn out to be complex.

The fixed-dimension perturbative expansion in powers of two independent
zero-momentum quartic couplings $u$ and $v$ (directly related to 
$u_0$ and $v_0$ defined in Eq.~(\ref{Hphi4rim})) is apparently better
behaved. Up to six loops, a random fixed point $u^*$, $v^*$ is always 
found, although the estimates of $u^*$ and $v^*$ vary
significantly with the order and the resummation method.
In spite of that, the estimates of the critical exponents are 
quite precise, due to a relatively large insensitivity
of the results to the position of the fixed point.
The analysis of the six-loop series gives \cite{PV-00} 
$\nu=0.678(10)$, $\eta=0.030(3)$, $\beta=0.349(5)$, and $\alpha=-0.034(30)$.
The agreement with the experimental and numerical results
is again quite satisfactory; only the estimate of $\alpha$
seems to be slightly larger than the experimental result.

In this paper we present a new numerical study 
of the RIM. The purpose is to extend and possibly 
improve the numerical results of Ref.~\cite{BFMMPR-98}.
We estimate the critical exponents and, in particular, $\alpha$ in order to 
verify if the apparent small discrepancy between experiments and 
numerical results is really there. Moreover,
we determine the four-point and the six-point zero-momentum couplings, 
and the universal ratio $R_\xi^+$. As a byproduct we are able to check the 
accuracy of the FT approach by comparing Monte Carlo and FT estimates of 
the fixed-point values $u^*$ and $v^*$. 

We have performed a high-precision Monte Carlo simulation of the model
with Hamiltonian (\ref{latticeH}) at $J=1$ and density 
$x = 0.8$. Such a value has been chosen on the basis of the results 
of Ref.~\cite{BFMMPR-98}, where it was shown that 
scaling corrections are particularly small for such a value of $x$. 
This is fully confirmed 
by our analysis: We do not observe scaling corrections with exponent
$\omega = 0.37(6)$, the correction-to-scaling 
exponent observed in Ref.~\cite{BFMMPR-98} for generic values of the 
density.\cite{foot-omega-FT} Note that the absence of corrections with
exponent $\omega$ also implies the absence of corrections with 
exponents $2\omega$, $3\omega$, $\ldots$ Therefore, we expect corrections
to scaling with next-to-leading exponent $\omega_2$ 
($\omega_2 = 0.8(2)$  according to field theory \cite{CPPV-03}). 
Unexpectedly, also these corrections are small.
The RIM at $x = 0.8$ is therefore an ``improved" model
\cite{CFN-82,BFMM-98,HPV-99,CHPRV-02,CPRV-02}, 
i.e. a model in which the leading correction to scaling is 
(approximately) absent in the expansion of any observable near the critical 
point. 

First of all, we determine the critical exponents
by using two different methods. A first estimate is
obtained by employing the extrapolation method of 
Refs.~\cite{CEFPS-95,CEPS-95,MPS-97} (similar methods have been
discussed in Refs.~\cite{LWW-91,Kim-94}). It allows us to determine 
the critical exponents from the high-temperature behavior of the 
susceptibility and of the correlation length. We also use direct 
finite-size scaling (FSS) methods, obtaining consistent estimates. 
Our final results are
\begin{eqnarray}
&& \nu = 0.683(3), \label{stima-nu-sez1} \\
&& \eta = 0.035(2),
\end{eqnarray} 
from which, using scaling and hyperscaling relations, we obtain
\begin{eqnarray}
&& \gamma = \nu (2 - \eta) = 1.342(6), \\
&& \beta = {\nu\over2}(1 + \eta) = 0.3535(17), \\
&& \delta = {5 - \eta\over 1 + \eta} = 4.80(11), \\
&& \alpha = 2 - 3 \nu = -0.049(9).
\end{eqnarray} 
Our results are in good agreement with those of Ref.~\cite{BFMMPR-98}
and, in particular, confirm the discrepancy between the 
experimental and theoretical estimates of $\alpha$.

We also carefully check the validity of the hyperscaling relation 
$2 - \alpha = 3 \nu$. We analyze the specific heat at the critical point
obtaining $\alpha/\nu = -0.115(28)$.
Using the estimate (\ref{stima-nu-sez1}) for $1/\nu$ we obtain 
\begin{equation} 
 {2\over \nu} - {\alpha\over \nu} = 3.04(3),
\end{equation}
which is fully consistent with 3. We also perform another check of 
hyperscaling, analyzing the specific heat and the energy at the critical point.
We obtain 
\begin{equation}
 {2\over \nu} - {\alpha\over \nu} = 2.93(6),
\end{equation}
again consistent with 3.

Beside the critical exponents we also measure the four-point couplings 
$G^*_4$ and $G^*_{22}$ defined in Eqs.~(\ref{fourpoint-def}) and 
(\ref{limSstar}), which can be 
directly related to the fixed-point values $u^*$ and $v^*$: 
$G^*_4 = v^*$ and $G^*_{22} = u^*/3$. We obtain:
\begin{eqnarray}
&& G^*_4 = 43.3(2), \nonumber \\
&& G^*_{22} = -6.2(1) \; .
\end{eqnarray}
These estimates differ significantly from those reported in Ref.~\cite{PV-00},
which were obtained from the analysis of perturbative six-loop series:
$G^*_4 = 38.0(1.5)$ and $G^*_{22} = -4.5(6)$. Clearly, the non-Borel 
summability of the perturbative expansions 
gives rise to a large systematic error.
It is also possible that the nonanalyticity of the RG functions 
\cite{Nickel-81,PV-98,CCCPV-00,CPV-Kleinert}
near the random fixed point plays an important role.

These discrepancies on the estimates of $u^*$ and $v^*$ call for a reanalysis 
of the perturbative expansions of the critical exponents. 
By using the Monte Carlo estimate of $u^*$ and $v^*$ we find 
$\nu = 0.686(4)$, $\eta = 0.026(3)$, and $\gamma = 1.355(8)$. 
These estimates do not differ significantly from those 
obtained in Ref.~\cite{PV-00} and are also in satisfactory agreement 
with the Monte Carlo results. Clearly exponents are quite insensitive to the 
exact location of the fixed point.
We also try a different method for estimating critical 
quantities. It is based on an expansion around the Ising fixed point. 
Results are similar: $\nu = 0.690(8)$, $\eta = 0.0345(20)$, and 
$\gamma = 1.355(10)$. Note that the estimate of $\eta$ is now in perfect
agreement with the Monte Carlo result. 

In this paper we also determine some other universal amplitude ratios
that involve high-temperature quantities. First, we determine the six-point
universal ratios $r_6^*$, $C_{42}^*$, and $C_{222}^*$, defined in 
Eq.~(\ref{sixpoint-def}). The coefficient $r_6^*$ is particularly important 
since it parametrizes the small-magnetization expansion of the 
critical equation of state in the high-temperature phase 
\cite{CPV-03-eqstate}. We find 
\begin{equation}
 r_6^*  = 0.90(15).
\end{equation}
Finally, we compute the universal ratio $R^+_\xi$ defined by
\begin{equation}
R^+_\xi \equiv (\alpha A^+)^{1/3} f^+,
\end{equation}
where $A^+$ and $f^+$ are defined in terms of the singular behavior 
of the specific heat $C$ and of the correlation length $\xi$ ,
$C_{\rm sing} \approx  A^+ t^{-\alpha}$, 
$\xi \approx f^+ t^{-\nu}$ for $t\equiv \beta_c - \beta \to 0^+$. We obtain 
\begin{equation}
   R^+_\xi = 0.2885(15),
\end{equation}
in good agreement with other theoretical results \cite{CPV-03-eqstate}:
$R^+_\xi = 0.290(10)$, obtained from the analysis of the corresponding 
six-loop perturbative series, and $R^+_\xi = 0.282(3)$, 
derived from a quite precise approximation of the equation of state.

The paper is organized as follows. 
In Sec. \ref{secII} we present the Monte Carlo results. 
In Sec.~\ref{secIIB} we determine the critical temperature by performing 
a careful analysis of the finite-size behavior of some 
RG invariant ratios near 
the critical point. In Sec.~\ref{secIIC} we determine the four-point and 
six-point couplings by using the extrapolation method 
of Refs.~\cite{CEFPS-95,CEPS-95,MPS-97}. In Sec.~\ref{secIID} and
\ref{secIIE} we determine $\nu$ and $\eta$ by using again the extrapolation
method and by also performing a more direct FSS analysis.
Then, in Sec.~\ref{secIIF} we study the finite-size behavior of the 
energy and of the specific heat near the critical point. 
We obtain an independent estimate of $\alpha$, that allows us to check the 
validity of the hyperscaling relation $2 - \alpha = 3\nu$. Finally,
in Sec.~\ref{secIIG} we compute the universal ratio $R_\xi^+$. 
For this purpose, we generalize the extrapolation method of 
Refs.~\cite{CEFPS-95,CEPS-95,MPS-97} to the energy. In spite of the 
necessary subtractions, the method works quite well, providing a rather 
precise estimate. In Sec.~\ref{fieldtheory} we reanalyze the 
six-loop perturbative series of Ref.~\cite{PV-00}, using the 
new Monte Carlo estimate of the fixed point. We employ the 
different resummation methods discussed in Ref.~\cite{PV-00} and also 
a new method based on an expansion around the Ising fixed point.
Finally, in the appendix we report the definitions of the quantities
that are used throughout the paper.

\section{Numerical results} \label{secII}

\subsection{The Monte Carlo simulation} \label{secIIA}

We have performed a high-precision Monte Carlo simulation of the model
with Hamiltonian (\ref{latticeH}) with $J=1$ at density 
$x = 0.8$. Such a value has been chosen on the basis of the results 
of Ref.~\cite{BFMMPR-98}, who showed that for such a value of $x$ 
scaling corrections are particularly small. 
In the simulations we have considered cubic lattices of size $L^3$, 
$L=16$, 32, 64, 128, and 256, with periodic boundary conditions. 
Simulations have been performed 
for several values of $\beta$ between 0.275 and $0.28578$. 
Two thirds of the simulations refer to the interval 
$0.275 \le \beta \le 0.2856$ (we will call the corresponding data the 
high-temperature results), while one third of the CPU time
was used in simulations in a narrow interval around the critical point, 
$0.28572 \le \beta \le 0.28578$. The average number of samples
for each $\beta$ and $L$ has been approximately
$7\cdot 10^4$ ($L=16$), $36\cdot 10^3$ ($L=32$), $27\cdot 10^3$ ($L=64$), 
$12\cdot 10^3$ ($L=128$), and $3\cdot 10^3$ ($L=256$).
The runs were performed on 
a cluster with Dual Athlon MP 1.2MHz processors. The total CPU time is
approximately 17.4 CPU-years of a single processor.
As random number generator we have used a combination of the Parisi-Rapuano
generator \cite{PR-85} and of a congruential generator \cite{foot-random}.
Results for each sample have been obtained as follows. 
Starting from a random spin
configuration, we perform 2000 iterations, each of them consisting 
alternatively of a Metropolis sweep and of a full Swendsen-Wang 
update. We use both a local and a nonlocal dynamics to guarantee 
equilibration of short-distance and long-distance modes. 
Then, we perform 2000 full Swendsen-Wang
updates, measuring all quantities (see appendix for 
definitions) every 10 iterations. 
To estimate correlation functions we use improved estimators that 
significantly reduce the statistical errors. Note that we have been much more 
conservative than Ref.~\cite{BFMMPR-98}: There, only 200 
Swendsen-Wang iterations were performed for equilibration.
For quantities that involve the noise average of products 
of sample averages, there is a bias due to the finite length of the run
for each sample. In order to take this bias into account we have performed a 
bias correction following Ref.~\cite{BFMMPR-98-2}. 

\subsection{Determination of the critical temperature} \label{secIIB}

As a first step in our analysis we have determined the critical 
temperature $\beta_c$. For this purpose we consider the results of 
the simulations for $0.28572\le \beta \le 0.28578$, which is a small interval
around $\beta_c$. We consider four invariant 
ratios, $U_4$, $U_6$, $U_{22}$, and $\xi/L$, 
see appendix for definitions. Standard FSS predicts that, 
in the FSS limit $\beta\to \beta_c$, $L\to \infty$ at
$(\beta - \beta_c) L^{1/\nu}$ fixed,
each quantity $R(\beta,L)$ behaves as 
\begin{equation}
   R(\beta,L) = \hat{R}[(\beta - \beta_c) L^{1/\nu}],
\end{equation}
where $\hat{R}(z)$ is a universal function.
Since $\beta-\beta_c$ is particularly 
small for the data, we can expand $\hat{R}(z)$ in powers of $z$, keeping 
only the first term (we checked that the addition of the term of order 
$z^2$ does not change the results).
Thus, we fit each quantity $R(\beta,L)$ by using 
\begin{equation}
  R(\beta,L) = R^* + a (\beta - \beta_c) L^{1/\nu},
\end{equation}
with $R^*$, $\beta_c$, and $\nu$ free parameters. 
In each fit we include all data with $L\ge L_{\rm min}$ and,
in order to detect corrections to scaling, we use
$L_{\rm min} = 16$, 32, and 64.
The results are reported in Table \ref{Binder-betac}.

\begin{table}
\begin{center}
\caption{Results of the fit $R(\beta,L) = R^* + a (\beta - \beta_c) L^{1/\nu}$. 
DOF is the number of degrees of freedom of the fit.}
\label{Binder-betac}
\begin{tabular}{rrlll}
$L_{\rm min}$ & $\chi^2$/DOF &
     \multicolumn{1}{c}{$R^*$} &
     \multicolumn{1}{c}{$\beta_c$} &
     \multicolumn{1}{c}{$\nu$}  \\
\hline
\multicolumn{5}{c}{$U_4$} \\
  16 & 115.2/25 &  1.6337(4) & 0.2857520(7) &  0.804(44) \\
  32 &  17.5/20 &  1.6385(6) & 0.2857477(6) &  0.727(43) \\
  64 &   7.4/12 &  1.6407(11) & 0.2857462(9) &  0.726(60) \\
\multicolumn{5}{c}{$U_6$} \\
  16 & 134.4/25 &  3.2348(22) & 0.2857530(8) &  0.817(46) \\
  32 &  18.8/20 &  3.2597(28) & 0.2857480(6) &  0.727(43) \\
  64 &   7.4/12 &  3.2710(54) & 0.2857463(9) &  0.726(60) \\
\multicolumn{5}{c}{$U_{22}$} \\
  16 &  35.0/25 &  0.1500(8) & 0.2857266(141) &  1.23(51) \\
  32 &  25.0/20 &  0.1484(6) & 0.2857425(68) &  1.06(54) \\
  64 &  21.2/12 &  0.1480(10) & 0.2857454(83) &  0.93(66) \\
\multicolumn{5}{c}{$\xi/L$} \\
  16 &  22.6/25 &  0.5921(2) & 0.2857414(6) &  0.733(34) \\
  32 &  17.4/20 &  0.5926(3) & 0.2857423(6) &  0.708(37) \\
  64 &  11.8/12 &  0.5934(6) & 0.2857432(9) &  0.722(54) \\
\end{tabular}
\end{center}
\end{table} 

There is a slight evidence of corrections to scaling, but it is interesting to
note that they have opposite sign in $U_4$, $U_6$, and $\xi/L$. 
Conservatively, we would obtain
\begin{equation}
\beta_c = 0.2857447(24),
\label{betac-conservative}
\end{equation}
where the central value is the average of the estimates obtained for
$U_4$ and $\xi/L$ ($L_{\rm min} = 64$), and 
the error is such to include one error bar
for both $U_4$ and $\xi/L$. 

These fits are not particularly sensitive to the exponent $\nu$, which is 
quite poorly determined. We obtain $\nu = 0.72(6)$.
One could imagine of improving the results by fixing
$\nu$. However, the dependence of $\beta_c$ on $\nu$ is very small and
no significant change is observed.

In order to include scaling corrections we also perform
fits of the form
\begin{equation}
  R(\beta,L) = R^* + a (\beta - \beta_c) L^{1/\nu} + b L^{-\omega},
\label{fitR-corr}
\end{equation}
where we include the leading scaling correction. These fits are not 
sensitive to the value of $\nu$ and thus we fix it, taking $\nu = 0.69$. 
We keep $\omega$ as a free parameter,
since we do not know which is the most important correction to scaling for 
our data. Indeed, the leading correction has exponent $\omega = 0.37(6)$, 
but there is evidence that for $p=0.8$ leading corrections have a 
very small amplitude \cite{BFMMPR-98}. In order to be able to keep 
$\omega$ as a free parameter, we analyzed at the same time two different 
observables. We restrict our attention to $U_4$, $U_6$, and $\xi/L$,
since $U_{22}$ is too noisy.
Using all data with $L\ge 16$ (if the data with $L=16$ are discarded
the fit is unstable) and taking properly into account the covariance
between the two observables, we obtain:
\begin{itemize}
\item[(a)]
 analysis of $\xi/L$ and $U_4$:
$\omega = 0.70(11)$, $\beta_c = 0.2857435(8)$,
$(\xi/L)^* = 0.5943(8)$, $U_4^* = 1.6502(24)$, $b(\xi/L) = -0.017(7)$,
$b(U_4) = -0.13(3)$; $\chi^2 = 47.5$, DOF$=50$.
 
\item[(b)]
analysis of $\xi/L$ and $U_6$:
$\omega = 0.71(11)$, $\beta_c = 0.2857435(8)$,
$(\xi/L)^* = 0.5942(8)$, $U_6^* = 3.318(11)$, $b(\xi/L) = -0.017(7)$,
$b(U_4) = -0.67(16)$; $\chi^2 = 46.7$, DOF$=50$.
\end{itemize}
Here DOF is the number of degrees of freedom.
The results are quite stable, indicating the presence of corrections
with exponent $\omega = 0.7(1)$, 
in agreement with the idea that scaling corrections with  
exponent $\omega \approx 0.4$ are very small. The effective exponent 
$\omega = 0.7(1)$ is quite close to the next-to-leading exponent 
predicted by perturbative field theory \cite{CPPV-03}, 
i.e. $\omega_2 = 0.8(2)$.
Thus, we mainly observe next-to-leading corrections, which in any 
case are quite small.
In particular, they are of little relevance for $\xi/L$.
The coefficient $b$, cf.  Eq.~(\ref{fitR-corr}), is very small and
the estimates of $\beta_c$ obtained from 
the combined fits are fully compatible with those obtained for 
$\xi/L$ without scaling corrections.

These analyses that keep into account scaling corrections hint at
values of $\beta_c$ lower than the estimate (\ref{betac-conservative}).
We are thus led to consider
\begin{equation}
    \beta_c = 0.285744(2)
\label{betac-final}
\end{equation}
as our final estimate.

From the above-reported analyses, we also obtain estimates of the 
invariant ratios $R^*$ at the critical point. We obtain
\begin{eqnarray}
\left({\xi\over L}\right)^* &=& 0.5943(9), \nonumber \\
U_4^* &=& 1.650(9), \nonumber \\
U_6^* &=& 3.32(5), \nonumber \\
U_{22}^* &=& 0.1480(10).
\end{eqnarray} 
We quote the results obtained in the fits (\ref{fitR-corr}), 
while the error is such to include also the result of the fit 
without scaling corrections and $L_{\rm min} = 64$. For $U_{22}$ we only 
consider the fits without scaling corrections. Note that $U_{22}^*\not=0$, 
indicating the absence of self-averaging at the critical point, in 
agreement with the theoretical arguments of Ref.~\cite{AH-96}.

We can compare our results with those of Ref.~\cite{BFMMPR-98}. They found
$\beta_c = 0.2857421(52)$, 
$U_4^* = 1.653(20)$,
$U_{22}^* = 0.145(7)$,
$(\xi/L)^* = 0.598(7)$, 
which are in full agreement with our final results. 

\subsection{Determination of the four-point and 
six-point couplings} \label{secIIC}

In this Section we wish to determine the four-point couplings 
$G_4^*$ and $G_{22}^*$, and the six-point universal ratios $r_6^*$,
$C_{42}^*$, and $C_{222}^*$, see appendix for definitions. 
Note that these quantities are defined in the 
high-temperature phase and one should take first the 
infinite-volume limit and then the limit $\beta\to \beta_c$,
cf. Eq.~(\ref{limSstar}).
In order to perform this task we have applied the extrapolation method 
of Refs.~\cite{CEFPS-95,CEPS-95,MPS-97} to our 
high-temperature data, i.e. to the results with 
$\beta \le 0.2856$ (corresponding to $\xi_\infty(\beta) \lesssim 89$). 
This method is extremely powerful 
in order to compute the infinite-volume behavior of critical quantities
and it has been applied to several models, including spin glasses \cite{PC-99}.

The idea is the following. Given a long-distance quantity 
$S(\beta,L)$, in the FSS limit we can write
\begin{equation}
{S(\beta,sL)\over S(\beta,L)} =
   F_S\left[\xi(\beta,L)/L\right] + O(L^{-\omega},\xi^{-\omega}),
\label{stl3}
\end{equation}
where $s$ is an arbitrary (rational) number 
(here we always consider $s=2$). Here $F_S(z)$ is a universal function
defined for $0\le z \le z^* \equiv (\xi/L)^* = 0.5943(9)$ 
such that $F_S(0) = 1$ and $F_S(z^*) = s^\sigma$, where 
$\sigma$ is the exponent characterizing the critical behavior of 
$S(\beta,L)$ at the critical point, i.e., $S(\beta_c,L)\sim L^\sigma$. 
Eq.~(\ref{stl3}) is the basis of the extrapolation technique since, in 
the absence of scaling corrections, it allows us to compute 
$S(\beta,sL)$ on a lattice of size $sL$ in terms of quantities defined 
on a lattice of size $L$ and of the function $F_S(z)$. In practice,
one works as follows.
First, one performs several runs, determining
$S(\beta,sL)$, $S(\beta,L)$, $\xi(\beta,sL)$, and $\xi(\beta,L)$.
By means of a suitable interpolation, this provides the
function $F(z)$ for $S$ and $\xi$.  Then,
$S_\infty(\beta)$ and $\xi_\infty(\beta)$ are obtained
from $S(\beta,L)$ and $\xi(\beta,L)$
by iterating Eq.~(\ref{stl3}) and the corresponding equation for
$\xi(\beta,L)$. Of course, one must be very careful about scaling
corrections, discarding systematically lattices with small values of $L$ 
till the results become independent of $L$ within error bars.  

\begin{figure}[t]
\centerline{\psfig{width=8truecm,angle=-90,file=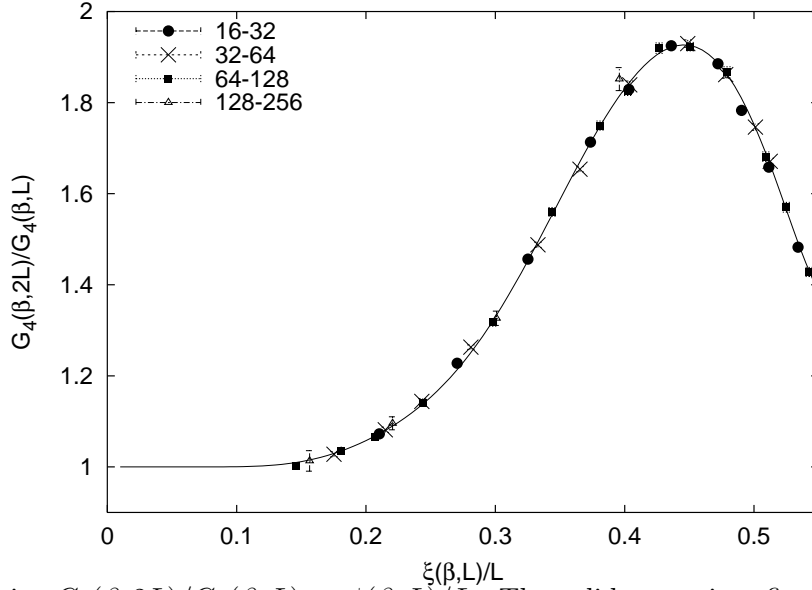}}
\caption{Ratios $G_4(\beta,2L)/G_4(\beta,L)$ vs $\xi(\beta,L)/L$.
The solid curve is a fit using all data with $L\ge 64$.
}
\label{figFSSG4}
\end{figure} 

\begin{figure}[t]
\centerline{\psfig{width=8truecm,angle=-90,file=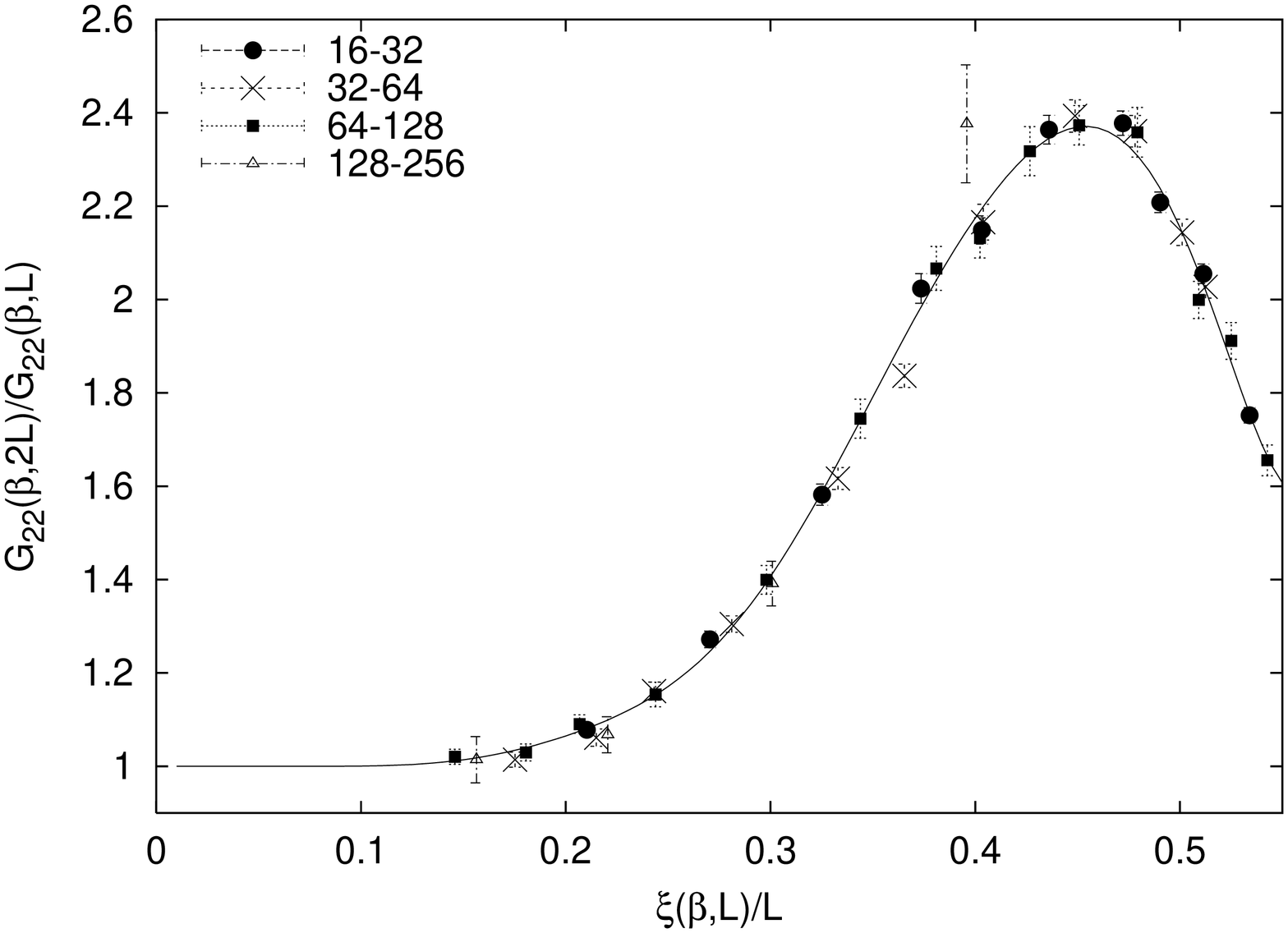}}
\caption{Ratios $G_{22}(\beta,2L)/G_{22}(\beta,L)$ vs $\xi(\beta,L)/L$.
The solid curve is a fit using all data with $L\ge 64$.
}
\label{figFSSG22}
\end{figure} 

\begin{figure}[t]
\centerline{\psfig{width=8truecm,angle=-90,file=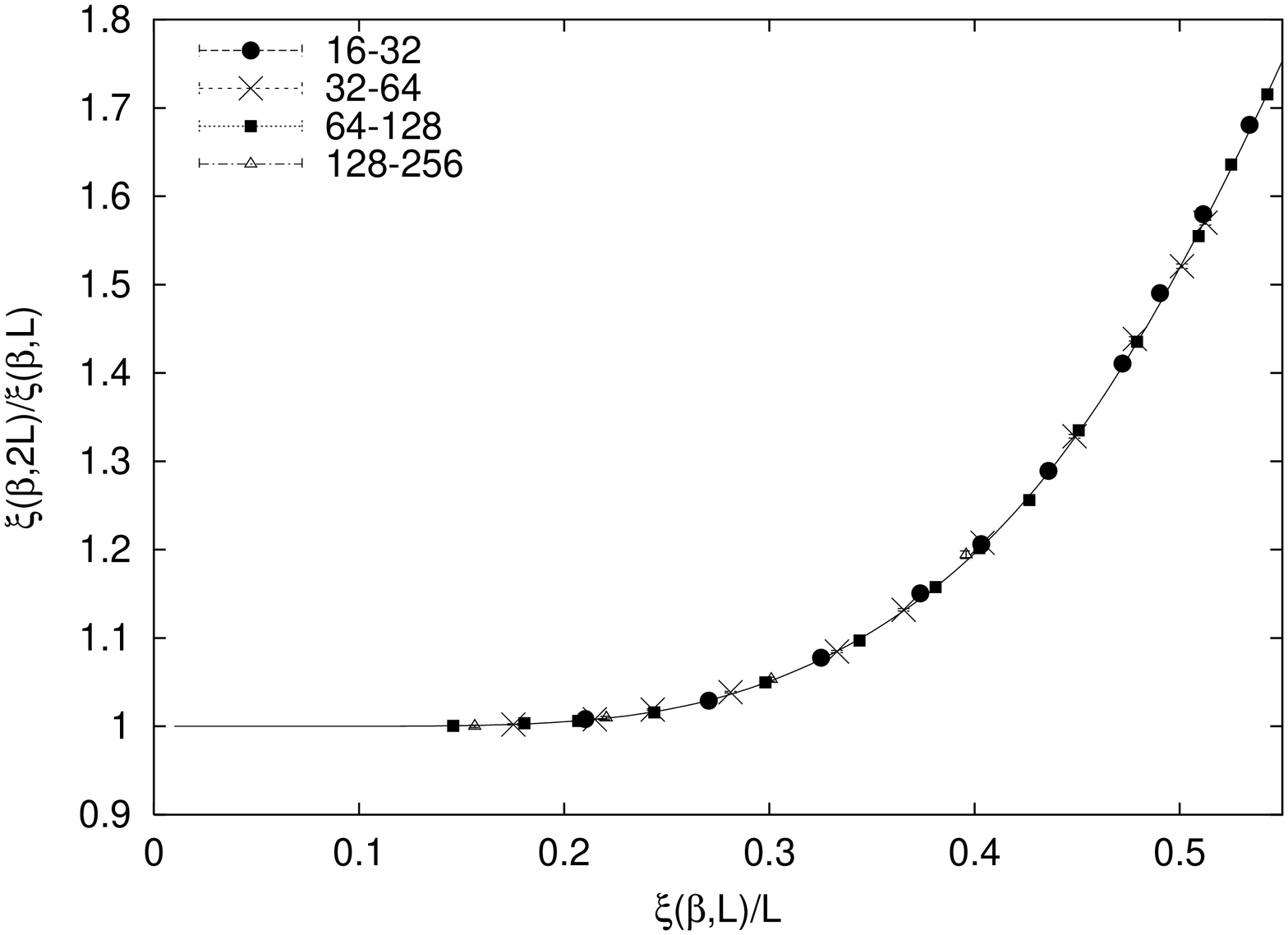}}
\caption{Ratios $\xi(\beta,2L)/\xi(\beta,L)$ vs $\xi(\beta,L)/L$.
The solid curve is a fit using all data with $L\ge 64$.
}
\label{figFSSxi}
\end{figure} 

Let us first discuss the four-point couplings for which we will 
obtain quite precise estimates. In Figs.~\ref{figFSSG4}, \ref{figFSSG22}, 
and \ref{figFSSxi} we report the data for $S(\beta,2L)/S(\beta,L)$ 
for $G_4$, $G_{22}$, and $\xi$, together with a fit of the data with 
$L\ge 64$. As discussed in Refs.~\cite{CEFPS-95,CEPS-95}, we parametrize 
$F_S(z)$ with a polynomial \cite{footJacobi} in $e^{-1/z}$ of order $n$,
increasing $n$ until the $\chi^2$ of the fit changes by less than one by 
going from $n$ to $n+1$. In these analyses we have taken $n=7$. 
The parametrization of $F_S(z)$ as a polynomial in $e^{-1/z}$
is theoretically motivated: indeed, for zero-momentum quantities, 
$F_S(z)$ approaches 1 with corrections of order $e^{-a/z}$, 
$a \approx 1$, as $z\to 0$. 
This choice is not strictly correct for $\xi$ since in this case
\cite{CP-98} $F_\xi(z) = 1 + O(z^2)$. However, these power corrections 
are expected to be very small for our definition of finite-volume
correlation length \cite{foot-corrlength}, and therefore the systematic error 
due to our choice of parametrization should be small.

Looking at the figures, it is quite difficult to distinguish any correction 
to scaling, i.e. systematic deviations from the fitted curve. 
However, at a closer look one may see that some points with $L=16$ 
are out of the curve (in all cases by less than three error bars, so that 
these differences are barely significant). 
Conservatively, we have decided to discard 
all $L=16$ data. In order to check further for corrections to scaling 
we have computed infinite-volume estimates $S_\infty(L)$ using 
only data with $L\ge L_{\rm min}$, $L_{\rm min} = 32$, 64. Additionally,
we have also systematically discarded points that are far from the critical 
point by including only data with $\beta\ge \beta_{\rm min}$ 
for several values of $\beta_{\rm min}$. 

\begin{table}
\caption{Results for the renormalization constant $G_4^*$. 
On the left we report
the results for $L_{\rm min} = 32$, on the right those for 
$L_{\rm min} = 64$. We report two different $\chi^2$. 
The first one ($\chi^2_{\rm estr}$) refers to the fit that allows the 
determination of the curve $F_{G_4}(z)$, cf. Eq.~(\protect\ref{stl3}), 
the second one ($\chi^2_{\rm fit}$) to the fit 
(\protect\ref{extrap-to-betac}). Beside the $\chi^2$ we also report 
the number of degrees of freedom (DOF).
The results have two errors: the first one is the statistical error, 
the second one gives the variation of the estimate as $\beta_c$ is 
varied within one error bar, cf. Eq.~(\protect\ref{betac-final}).
}
\label{results-G4}
\begin{center}
\begin{tabular}{ccccccc}
$\beta_{\rm min}$ & $\chi^2_{\rm estr}$/DOF & $\chi^2_{\rm fit}$/DOF &
    $G_4^*$  & $\chi^2_{\rm estr}$/DOF & $\chi^2_{\rm fit}$/DOF &
    $G_4^*$ \\
\hline
 0.2750 & 14.9/22 & 17.3/16 &  43.65(6+0) &
         &  &   \\
 0.2780 & 14.9/22 & 16.8/15 &  43.63(6+0) &
        9.0/11 & 15.9/15 &  43.56(9+2) \\
 0.2800 & 14.9/21 & 13.4/14 &  43.59(7+0) &
        9.0/11 & 13.1/14 &  43.51(10+0) \\
 0.2810 & 14.9/20 & 11.9/13 &  43.60(8+1) &
        9.0/11 & 12.7/13 &  43.52(10+1) \\
 0.2820 & 14.8/19 &  9.2/12 &  43.55(8+1) &
        9.0/11 & 11.8/12 &  43.47(11+0) \\
 0.2830 & 12.1/17 &  9.1/11 &  43.41(11+1) &
        7.4/10 & 12.7/11 &  43.45(12+0) \\
 0.2835 & 12.0/15 &  7.6/10 &  43.34(12+1) &
        7.4/9 & 10.8/10 &  43.40(15+1) \\
 0.2840 &  9.1/13 &  9.3/9 &  43.34(17+2) &
        7.1/8 &  9.7/9 &  43.32(18+1) \\
\end{tabular}
\end{center}
\end{table}

\begin{table}
\caption{Results for the renormalization constant $G_{22}^*$. 
Definitions are as in Table \ref{results-G4}.  }
\label{results-G22}
\begin{center}
\begin{tabular}{ccccccc}
$\beta_{\rm min}$ & $\chi^2_{\rm estr}$/DOF & $\chi^2_{\rm fit}$/DOF &
    $G_{22}^*$  & $\chi^2_{\rm estr}$/DOF & $\chi^2_{\rm fit}$/DOF &
    $G_{22}^*$ \\
\hline
 0.2750 & 19.5/22 & 13.6/16 &  $-$6.18(3+0) &
        &  &  \\
 0.2780 & 19.5/22 & 14.2/15 &  $-$6.18(3+0) &
       12.0/11 & 11.3/15 &  $-$6.25(4+0) \\
 0.2800 & 19.1/21 & 13.1/14 &  $-$6.18(3+0) &
       12.0/11 & 11.5/14 &  $-$6.25(4+0) \\
 0.2810 & 17.4/20 & 10.9/13 &  $-$6.22(3+0) &
       12.0/11 & 11.3/13 &  $-$6.25(5+1) \\
 0.2820 & 17.4/19 &  6.2/12 &  $-$6.19(3+0) &
       12.0/11 &  8.9/12 &  $-$6.23(5+0) \\
 0.2830 & 16.9/17 &  5.7/11 &  $-$6.18(4+0) &
       11.7/10 &  8.4/11 &  $-$6.23(6+0) \\
 0.2835 & 16.8/15 &  4.9/10 &  $-$6.19(5+1) &
       11.3/9 &  8.2/10 &  $-$6.25(6+1) \\
 0.2840 & 12.1/13 &  2.5/9 &  $-$6.12(7+0) &
       10.4/8 &  1.5/9 &  $-$6.11(8+1) \\ 
\end{tabular}
\end{center}
\end{table}

Using the extrapolation procedure we have outlined above, for each 
$L_{\rm min}$ and $\beta_{\rm min}$ we obtain infinite-volume 
estimates $\xi_\infty(\beta)$, $G_{4,\infty}(\beta)$, and 
$G_{22,\infty}(\beta)$. To obtain the estimate at the critical point,
the extrapolated values for the coupling constants have been fitted by using 
\begin{equation}
 S_\infty(\beta) = S^* + a (\beta_c - \beta),
\label{extrap-to-betac}
\end{equation}
with $\beta_c = 0.285744(2)$. The results are reported in 
Tables \ref{results-G4} and \ref{results-G22} 
for $0.275\le \beta_{\rm min}\le 0.284$ (corresponding to 
$4.45\lesssim \xi_\infty \lesssim 15.86$). 

To check for corrections to scaling, we have also performed a different 
analysis. First we fit $S(\beta,2L)/S(\beta,L)$ taking into account 
a scaling correction with exponent $\omega$, i.e. assuming
\begin{equation}
{S(\beta,2L)\over S(\beta,L)} = 
   F_S\left(\xi(\beta,L)/L\right)  + 
   {1\over L^\omega} G_S\left(\xi(\beta,L)/L\right),
\label{S2L-L-corr}
\end{equation}
and use both functions $F_S(z)$ and $G_S(z)$ to perform the infinite-volume
extrapolation. Then, the infinite-volume results for the coupling
constants are fitted by using 
\begin{equation}
 S_\infty(\beta) = S^* + a (\beta_c - \beta)^{\omega\nu}.
\label{extrap-to-betac-corr}
\end{equation}
In order to perform the analysis we should fix the exponents $\omega$ and
$\nu$. We use $\nu = 0.69$, and repeat the analysis using $\omega = 0.8$
(the next-to-leading exponent predicted by field theory \cite{CPPV-03})
and $\omega = 0.4$, which is the leading exponent determined in 
Ref.~\cite{BFMMPR-98}. The results corresponding to $\omega = 0.8$
are reported in Table \ref{results-G4-G22-corr}. 
It is essential to include the results with $L=16$ in the analysis, 
otherwise the data do not show FSS corrections and the fit is unstable. 
Therefore, we cannot check the goodness of the Ansatz by discarding 
data with small $L$, i.e. present results for different values of 
$L_{\rm min}$ as done before.

\begin{table}
\caption{Results for $G_{4}^*$ and $G_{22}^*$ for fits with a 
correction-to-scaling exponent $\omega=0.8$. 
We report two different $\chi^2$. 
The first one ($\chi^2_{\rm estr}$) refers to the fit that allows the 
determination of the curve $F(z)$, cf. Eq.~(\protect\ref{S2L-L-corr}), 
the second one ($\chi^2_{\rm fit}$) to the fit 
(\protect\ref{extrap-to-betac-corr}). Beside the $\chi^2$ we also report 
the number of degrees of freedom (DOF). We only report the statistical error;
the error due to $\beta_c$ is negligible.
}
\label{results-G4-G22-corr}
\begin{center}
\begin{tabular}{ccccccc}
$\beta_{\rm min}$ & $\chi^2_{\rm estr}$/DOF & $\chi^2_{\rm fit}$/DOF &
    $G_4^*$  & $\chi^2_{\rm estr}$/DOF & $\chi^2_{\rm fit}$/DOF &
    $G_{22}^*$ \\
\hline
 0.2750 & 30.0/29 & 12.6/16 &  43.28(8)   &
       31.8/29 & 12.1/16 &  $-$6.15(3) \\
 0.2780 & 29.9/28 & 10.9/15 &  43.31(9)   &
       31.6/28 & 11.1/15 &  $-$6.16(4) \\
 0.2800 & 29.5/26 & 10.8/14 &  43.30(10)   &
       31.1/26 &  9.9/14 &  $-$6.16(4) \\
 0.2810 & 25.4/24 &  9.0/13 &  43.43(11)   &
       22.1/24 &  8.8/13 &  $-$6.26(5) \\ 
 0.2820 & 24.3/22 &  9.0/12 &  43.30(13)   &
       22.0/22 &  5.8/12 &  $-$6.21(6) \\  
 0.2830 & 21.3/19 &  6.4/11 &  43.37(17)   &
       21.1/19 &  3.6/11 &  $-$6.25(7) \\
 0.2835 & 13.4/16 &  7.1/10 &  43.53(22)   &
       18.1/16 &  3.2/10 &  $-$6.33(9) \\
 0.2840 & 10.6/13 &  7.9/9 &  43.26(29)    &
       12.3/13 &  2.7/9 &  $-$6.00(13) \\
\end{tabular}
\end{center}
\end{table}

Let us first discuss the results for $G_4^*$. The fits without corrections 
to scaling show a significant (at the level of the reported errors)
decrease as $\beta_{\rm min}$ is increased and also a slight dependence on
$L_{\rm min}$. Corrections to scaling are positive and the estimate decreases
with increasing $\beta_{\rm min}$, 
so that one only obtains an upper bound $G_4^*\lesssim 43.4$. 
On the other hand, the fit
with $\omega = 0.80$ gives results independent of $\beta_{\min}$ 
within error bars; moreover the $\chi^2$ of the fit 
(\ref{extrap-to-betac-corr}) is systematically lower than that of the 
fit (\ref{extrap-to-betac}). Clearly the data are very well fitted by 
assuming a correction-to-scaling exponent $\omega = 0.80$. 
For $\omega = 0.40$ 
the results strongly depend on $\beta_{\rm min}$, varying from 
$G^*_4 = 42.49(14)$ for $\beta_{\rm min} = 0.275$ to 
$G^*_4 = 43.47(40)$ for $\beta_{\rm min} = 0.2835$. Also, the $\chi^2$ 
is larger than the $\chi^2$ obtained using $\omega = 0.8$. There is 
therefore little evidence for such a small correction-to-scaling 
exponent, confirming again that for density $x$=0.8 the leading 
scaling corrections are very small. As final estimate we take
\begin{equation}
G^*_4 = 43.3(2),
\label{Gstar4-final}
\end{equation}
which is consistent with all results.

Let us now consider the results for $G_{22}^*$. The results of 
Table \ref{results-G22} show no dependence on $\beta_{\rm min}$
and a tiny dependence on $L_{\rm min}$ which could be of purely 
statistical origin. With the present error bars there is no evidence 
for nonanalytic scaling corrections and indeed the results obtained 
using $\omega = 0.80$ (see Table \ref{results-G4-G22-corr})
are perfectly consistent with those of the fits
with purely analytic corrections. Our final estimate is 
\begin{equation}
G^*_{22} = - 6.2(1).
\label{Gstar22-final}
\end{equation}
The error is rather conservative and is such to include all estimates.

We should note that our results are compatible with the FT
predictions of Ref.~\cite{CPPV-03}, where it is shown that in infinite volume 
$G_{22,\infty}(\beta)$ and $G_{4,\infty}(\beta)$ 
have scaling corrections with next-to-leading exponent 
$\omega = 0.8(2)$ of similar relative size.
In particular, the results of Ref.~\cite{CPPV-03} give
$a_{G_{22}} = - 0.23(10) a_{G_{4}}$ where $a$ is the coefficient
defined in Eq.~(\ref{extrap-to-betac-corr}).
From the fits we obtain instead $a_{G_{4}} = 20(5)$ and 
$a_{G_{22}} = - 1.5(1.0)$. The errors should be taken with caution:
They simply give the variation of the parameter $a$
with $\beta_{\rm min}$ for 
$0.275 \le \beta_{\rm min} \le 0.281$ (for larger values the statistical error 
becomes larger than $|a|$) and do not include any 
possible systematic effect. Assuming these values with their errors,
we estimate $a_{G_{22}} = -0.08(5) a_{G_{4}}$,
which is in reasonable agreement with the FT result.

As a check we have repeated the analysis for $H_4 \equiv G_4 + 3 G_{22}$. 
Since the procedure is nonlinear, this represents an important
consistency check. We obtain
\begin{equation}
H_4^* = 24.7(2),
\label{Hstar4-final}
\end{equation}
which is in full agreement with the estimates reported above.

\begin{figure}[t]
\centerline{\psfig{width=8truecm,angle=-90,file=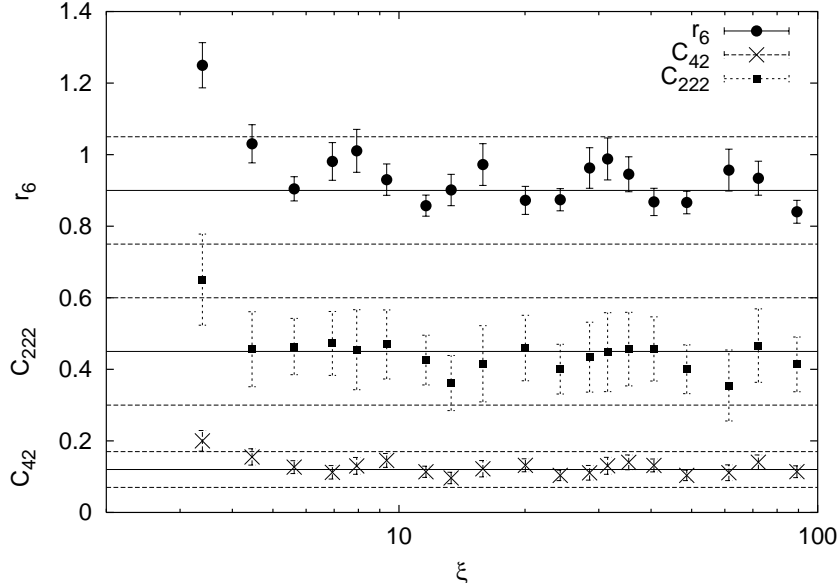}}
\caption{Infinite-volume results for the six-point ratios 
$r_{6}(\beta)$, $C_{42}(\beta)$, and $C_{222}(\beta)$ from the extrapolation 
with $L_{\rm min} = 16$, $\beta_{\rm min} = 0.275$, and degree of 
the interpolation polynomial $n = 7$. On the horizontal axis we report 
$\xi_\infty(\beta)$.
We also report the final results $r_6^* = 0.90(15)$, 
$C_{42}^* = 0.12(5)$, and $C_{222}^* = 0.45(15)$.
}
\label{fig-sixpoint}
\end{figure}

Finally, we consider the six-point couplings $r_6$, $C_{42}$, and $C_{222}$.
We apply again the extrapolation procedure we have used for $G_4$ and 
$G_{22}$. However, in this case there are larger systematic errors.
The extrapolation curve $F(z)$, cf. Eq.~(\ref{stl3}),
is poorly determined for $z \lesssim 0.3$, since
the six-point couplings have large statistical errors 
when $\xi(\beta,L)/L$ is small. A large error on the curve gives a 
large systematic error on the extrapolations and induces 
correlations among the results for different $\beta$ (such correlations
are instead small for $G_4$ and $G_{22}$). Because of them, it is  
difficult to set correct error bars and to determine the final results. 
For these reasons we have taken 
a conservative attitude. We have generated different sets of extrapolated data
by changing the degree $n$ of the interpolation polynomial and $\beta_{\min}$
(in all cases we set $L_{\rm min} = 16$ in order to have a sufficiently
large number of points in the small-$z$ region).  Then, we determine 
the range that includes most (approximately 2/3) of the extrapolated data 
with their error bars. The central value of such an interval gives the 
final result, while its half width gives the error. We obtain
\begin{eqnarray}
r_6^* &=& 0.90(15), \nonumber \\
C_{42}^* &=& 0.12(5), \nonumber \\
C_{222}^* &=& 0.45(15).
\label{risultati-sixpoint}
\end{eqnarray}
In Fig.~\ref{fig-sixpoint} we report the final results together with 
the results for a single extrapolation, so that the reader 
can see the quality of our numerical results and how much 
conservative our final error bars are.
Note that in this analysis we have not made any attempt to evaluate 
corrections to scaling. In any case, we expect them to be small 
compared with the large errors we quote.

The numerical estimates can be compared with the FT results. 
We shall discuss the four-point couplings in Sec.~\ref{fieldtheory}. 
For what concerns the six-point couplings, the analysis of 
the available four-loop series in the fixed-dimension scheme \cite{PSS-02} 
is not very precise. We only mention the estimate $r_6^* = 1.1^{+0.1}_{-0.5}$
reported in Ref.~\cite{CPV-03-eqstate}, which agrees with the 
more precise result (\ref{risultati-sixpoint}).

\subsection{The exponent $\nu$} \label{secIID}

In order to determine the exponent $\nu$ we have followed two different 
strategies: (a) we have determined extrapolated values $\xi_\infty(\beta)$
using our extrapolation method and then we have performed a fit 
$\xi_\infty(\beta) \sim (\beta_c-\beta)^{-\nu}$; 
(b) we have performed a purely FSS
analysis without extrapolations. The advantage of method (b) 
is that we can keep both $\beta_c$ and $\omega$ as free parameters, 
and thus check the previously determined value for $\beta_c$ 
and our claim that scaling corrections with small values of $\omega$ 
are tiny. 

\begin{figure}[t]
\centerline{\psfig{width=8truecm,angle=-90,file=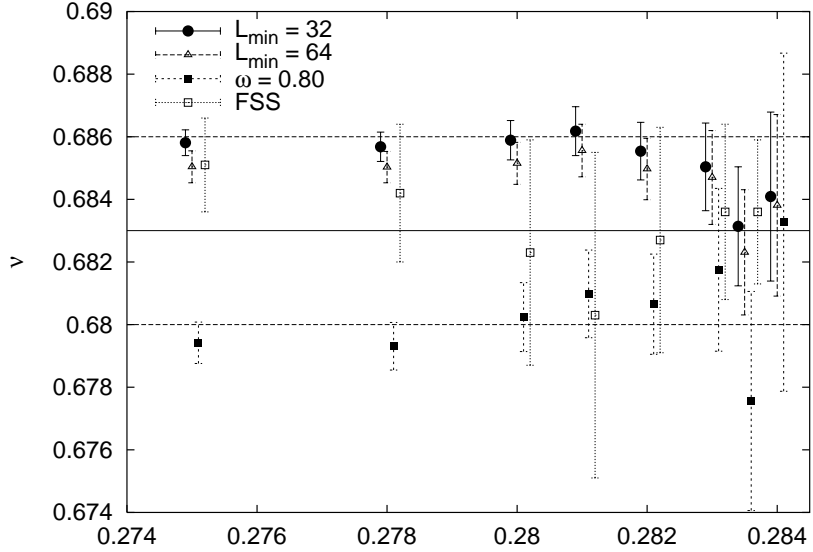}}
\caption{Results for the exponent $\nu$ vs $\beta_{\rm min}$.
We report the results obtained by fitting the extrapolated high-temperature
data without scaling corrections (they are labelled $L_{\rm min} = 32$ and 
$L_{\rm min} = 64$) and with scaling corrections with exponent $\omega = 0.8$
(labelled $\omega = 0.80$), and the results obtained by 
using all data and the parametrization (\ref{parametrization}) (labelled 
``FSS"). The horizontal lines correspond to the final result
$\nu = 0.683(3)$.
}
\label{fig-risnu}
\end{figure} 

In the first case we proceed as before. If we neglect scaling corrections, 
we can generate infinite-volume estimates $\xi_\infty(\beta)$
by using the extrapolation precedure based on Eq.~(\ref{stl3}) and 
then we can determine $\nu$ from fits of the form
\begin{equation}
\ln \xi_\infty(\beta) = - \nu\ln(\beta_c - \beta) + a + 
      b (\beta_c - \beta).
\label{fitxi-anal}
\end{equation}
As a second possibility we can include corrections with exponent $\omega$
in the extrapolation, see Eq.~(\ref{S2L-L-corr}), and fit the 
extrapolated data with 
\begin{equation}
\ln \xi_\infty(\beta) = - \nu\ln(\beta_c - \beta) + a +
      b (\beta_c - \beta)^{\omega\nu}.
\label{fitxi-nonanal}
\end{equation}
We report in Fig. \ref{fig-risnu}  the results 
for the analytic fit with $L_{\rm min} = 32$ and $L_{\rm min} = 64$ 
and for the nonanalytic fit with $\omega = 0.80$ and $L_{\rm min} = 16$ 
for several values of $\beta_{\rm min}$. 
We observe that results with and without nonanalytic corrections 
differ significantly, much more than the statistical errors.
However, corrections have opposite trends. 
Analytic fits give results that decrease as $\beta_{\rm min}$ is increased
while nonanalytic fits have the opposite behavior. 
Compatible results are obtained for $\beta_{\rm min} \ge 0.283$. 
For $\beta = 0.283$ we obtain $\nu = 0.6847(15)$ from the 
first analysis with $L_{\rm min} = 64$ and $\nu = 0.6818(24)$ from the 
second analysis. A reasonable estimate would be $0.683(3)$ which includes 
all results. It should be noted that the results depend strongly on
$\beta_c$ that has been fixed to $\beta_c = 0.285744(2)$. 
By varying $\beta_c$ within one error bar, $\nu$ varies approximately by 
$\pm 0.003$. Thus, collecting everything together, this method 
gives the final result
\begin{equation}
\nu = 0.683(3+3).
\label{nu-metodoa}
\end{equation}
The estimate (\ref{nu-metodoa}) 
has been obtained by using only the  high-temperature data,
i.e. those with $\beta \le 0.2856$. A more precise estimate can be 
obtained by performing a direct FSS analysis that allows us to use 
both the high-temperature and the critical-point results for the correlation
length. In this way, we do not need to fix $\beta_c$ and we can also 
keep $\omega$ as a free parameter. We start from the general FSS 
expression
\begin{eqnarray}
{\xi\over L} = F[u_1(t)L^{1/\nu}, u_2(t) L^{-\omega}],
\label{scaling-FSS-xisuL}
\end{eqnarray} 
where only one correction-to-scaling operator has been taken into account.
Here $t \equiv (\beta_c - \beta)$, and $u_1(t)$ and 
$u_2(t)$ are nonlinear scaling fields satisfying $u_1(0) = 0$ and 
$u_2(0) \not= 0$.
Equation (\ref{scaling-FSS-xisuL}) can be rewritten in the scaling limit as
\begin{eqnarray}
{\xi\over L} = f_1(t L^{1/\nu}) + L^{-1/\nu} f_2(t L^{1/\nu}) +
               L^{-\omega} f_3(t L^{1/\nu}),
\label{xisuL-FSS-gen}
\end{eqnarray}
with $f_1(0) \not= 0$ and $f_3(0) \not= 0$. 
The three scaling functions represent the three types of
contributions we expect: $f_1(x)$ is the leading FSS curve,
$f_2(x)$ corresponds to the analytic corrections,
and $f_3(x)$ is the nonanalytic FSS correction.
The function $f_2(x)$ is related to $f_1(x)$ by 
$f_2(x) \sim x^2 f_1'(x)$, a relation that follows from the fact that 
this correction is due to the expansion of the scaling field 
$u_1(t) = t + a t^2 + O(t^3)$.
The existence of the infinite-volume limit fixes the behavior of these
functions for $x\to\infty$: $f_1(x) \sim x^{-\nu}$, $f_2(x) \sim x^{1-\nu}$
and $f_3(x) \sim x^{(\omega - 1)\nu}$.
 
Now, we wish to determine $\beta_c$, $\nu$, and $\omega$ by fitting the 
numerical data for $\xi/L$ with Eq.~(\ref{xisuL-FSS-gen}). For this purpose,
we must parametrize the scaling functions
reported above with expressions containing a finite number of parameters.
Thus, we consider an even integer $n$ and parametrize:
\begin{eqnarray}
{\xi\over L} &=& \left[P_{n}(t L^{1/\nu}) +
            L^{-\omega} (1 + p t L^{1/\nu})^{\omega\nu}
            Q_{n} (t L^{1/\nu}) \right]^{- \nu/n}, \nonumber \\
P_{n}(x) &=& \sum_{i=0}^{n} a_i x^i, \qquad\qquad
Q_{n}(x) = \sum_{i=0}^{n/2} b_i x^{2i},
\label{parametrization}
\end{eqnarray}
where $t\equiv (\beta_c - \beta)$. This parametrization 
has the correct behavior both for small and large $x$ for any $n$. 
Note that our choice of writing $Q_{n}(x)$ as a polynomial in 
$x^2$ is due only to practical considerations: since corrections are small,
$Q_n(x)$ cannot be parametrized with as many parameters as $P_n(x)$;
on the other hand, to guarantee the correct large-$x$ behavior $Q_n(x)$ and 
$P_n(x)$ must have the same degree. We also made some analyses 
writing $Q_{n}(x)$ as a polynomial in $x^3$---in this case $n$ must be 
a multiple of 3---without finding significant 
differences.  The value of $n$ has been chosen on the basis of the 
$\chi^2$ of the fit: $n$ has been increased until the $\chi^2$ does not
change significantly as $n$ is increased by 2. The results of the 
analyses we present correspond to $n=12$.

In the analyses 
we have kept $\nu$, $\omega$, $p$, $\beta_c$, $\{a_i\}$, and $\{b_i\}$ as free
parameters. The fit is stable only if we include the data with $L= 16$, 
otherwise 
there are no scaling corrections at the level of our error bars and therefore
it is impossible to determine $\omega$. To check for corrections, we have 
systematically discarded points far from $\beta_c$, including in the 
analysis only data with $\beta\ge \beta_{\rm min}$ for several values of 
$\beta_{\rm min}$. 
In order to have a mostly linear fit, we have fitted
$({\xi\over L})^{-n/\nu}$. 
The results are reported in Table \ref{tab:nu-FSS}.

\begin{table}
\caption{Results of the FSS analysis using the parametrization
(\ref{parametrization}) with $n=12$ as a function of $\beta_{\rm min}$. 
Here DOF is the number of degrees of freedom of the fit. }
\label{tab:nu-FSS}
\begin{tabular}{llccc}
$\beta_{\rm min}$ & $\chi^2$/DOF & $\nu$ & $\omega$ & $\beta_c$ \\
\hline
0.2750 & 51/62 & 0.6851(15) & 1.71(18) & 0.2857434(7) \\
0.2780 & 50/60 & 0.6842(22) & 1.60(28) & 0.2857434(7) \\
0.2800 & 45/57 & 0.6823(36) & 1.32(35) & 0.2857433(7) \\
0.2810 & 43/54 & 0.6803(52) & 1.10(34) & 0.2857435(8) \\
0.2820 & 41/51 & 0.6827(36) & 1.51(66) & 0.2857433(7) \\
0.2830 & 39/47 & 0.6836(28) & 2.1(1.3) & 0.2857430(7) \\
0.2835 & 30/43 & 0.6836(23) & 3.6(2.5) & 0.2857428(8) \\
\end{tabular}
\end{table}

The results for $\beta_c$ are quite stable, in full agreement with the 
analyses reported in Sec. \ref{secIIB} for $\xi/L$ and with the 
final estimate (\ref{betac-final}). The estimates of $\omega$ 
vary significantly and have a quite large error. There is however 
very little evidence of scaling corrections with $\omega \lesssim 1$, 
as already discussed in Sec.~\ref{secIIB}. Finally, let us consider $\nu$.
As it can be seen from Fig.~\ref{fig-risnu} the results are in rough
agreement with those found before and apparently become independent of 
$\beta_{\rm min}$ for $\beta \gtrsim 0.282$. Our final estimate is 
\begin{equation}
\nu = 0.683(3),
\label{nu-final}
\end{equation}
which is compatible with the results of all analyses. 
The estimate (\ref{nu-final}) is in very good agreement with 
the result $\nu = 0.6837(53)$ of Ref.~\cite{BFMMPR-98}. 

\subsection{The exponent $\eta$} \label{secIIE}

In order to compute the exponent $\eta$ and correspondingly
$\gamma = (2-\eta)\nu$, one can analyze the critical behavior of the 
susceptibility $\chi$. However, we have found much more 
convenient to analyze $\zeta\equiv \chi/\xi^2$. Indeed, because of 
statistical correlations between $\chi$ and $\xi$, the relative error
on $\zeta$ is significantly smaller than that on $\chi$.
Moreover, the FT analysis of Ref.~\cite{CPPV-03}
indicates that $\zeta$ has much smaller scaling corrections than $\chi$.

\begin{figure}[t]
\centerline{\psfig{width=8truecm,angle=-90,file=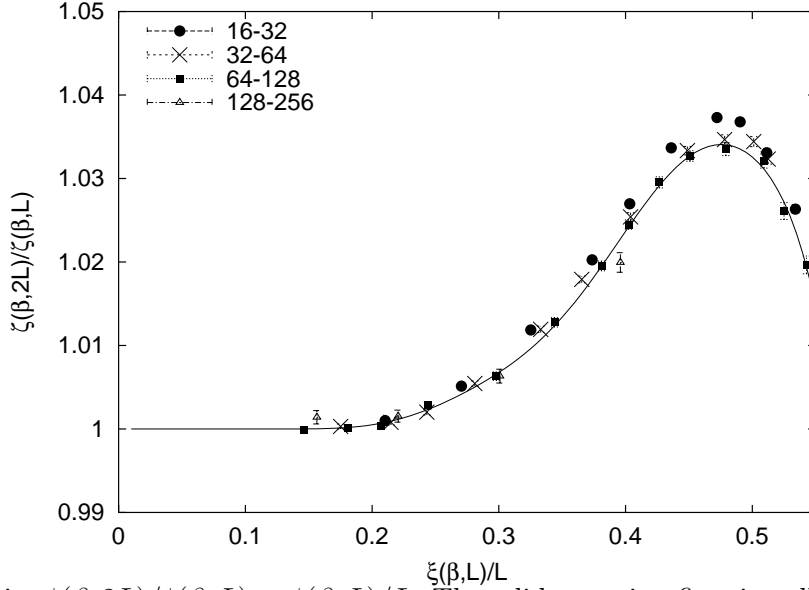}}
\caption{Ratios $\zeta(\beta,2L)/\zeta(\beta,L)$ vs $\xi(\beta,L)/L$.
The solid curve is a fit using all data with $L\ge 64$.
}
\label{figFSSZ}
\end{figure} 

In order to determine $\eta$ we perform two different analyses. 
The first one uses the high-temperature data and follows closely the 
analysis of the correlation length presented in Sec.~\ref{secIID}. 
Given $\zeta(\beta,L)$ we determine the curve $F_\zeta(z)$, 
cf. Eq.~(\ref{stl3}).
The ratios $\zeta(\beta,2L)/\zeta(\beta,L)$ are reported in Fig.~\ref{figFSSZ},
together with a fit of the points with $L\ge 64$. There are clear
scaling corrections, especially when $\xi(\beta,L)/L\gtrsim 0.3$: 
the data with $L=16$, and also some points with $L=32$, are systematically
above the curve. As before, we discard all points with $L=16$ and 
perform the extrapolation using data with $L\ge L_{\rm min}$, with 
$L_{\rm min} = 32,64$, in order to detect scaling corrections. The 
extrapolated values are fitted by using 
\begin{equation}
\ln \zeta_\infty (\beta) = 
\eta \nu \ln (\beta_c - \beta) + a + b (\beta_c - \beta)
\label{fitzeta-anal}
\end{equation}
and fixing $\beta_c = 0.285744(2)$. The results are reported in 
Table \ref{results-etanu}. They show a systematic dependence on 
$L_{\rm min}$ and $\beta_{\rm min}$ and seem to indicate 
$\eta\nu \approx 0.0240$, but it is difficult to set an error bar.
Conservatively, we write 
\begin{equation}
\eta\nu = 0.0240(15),
\label{stima-etanu}
\end{equation}
that includes the estimates with $\beta_{\rm min} = 0.284$ with their
error bars. Using the estimate of $\nu$ reported in Sec.~\ref{secIID},
$\nu = 0.683(3)$, we obtain  
\begin{equation}
\eta = 0.035(2).
\label{eta-final}
\end{equation}
We have also repeated the analysis including scaling corrections with 
$\omega = 0.8$. In the extrapolation procedure 
we use Eq.~(\ref{S2L-L-corr}), and then we fit the extrapolated data 
by using 
\begin{equation}
\ln \zeta_\infty (\beta) = \eta \nu \ln (\beta_c - \beta) + a + b 
    (\beta_c - \beta)^{\omega\nu}.
\label{fit-Z-omega}
\end{equation}
The results are reported in Table~\ref{results-etanu-corr}.
They show a systematic upward trend, indicating that scaling corrections with 
$\omega = 0.8$ do not correctly describe the data. This is also 
evident from the $\chi^2$ of the fit (\ref{fit-Z-omega}) which is 
quite large for small $\beta_{\rm min}$. 
A lower value of $\omega$, say $\omega = 0.4$, gives even worse results, 
confirming that corrections with $\omega\lesssim 1$ are very small,
in agreement with the FT analysis of Ref.~\cite{CPPV-03}.
In any case the results of Table~\ref{results-etanu-corr} indicate that 
$\eta \nu \gtrsim 0.0220$, in agreement with Eq.~(\ref{stima-etanu}). 

\begin{table}
\caption{Results for the critical exponent $\eta\nu$.
On the left we report
the results for $L_{\rm min} = 32$, on the right those for 
$L_{\rm min} = 64$. We report two different $\chi^2$. 
The first one ($\chi^2_{\rm estr}$) refers to the fit that allows the 
determination of the curve $F_\zeta(z)$, cf. Eq.~(\protect\ref{stl3}), 
the second one ($\chi^2_{\rm fit}$) to the fit 
(\protect\ref{fitzeta-anal}). Beside the $\chi^2$ we also report 
the number of degrees of freedom (DOF).
The results have two errors: the first one is the statistical error, 
the second one gives the variation of the estimate as $\beta_c$ is 
varied within one error bar, cf. Eq.~(\protect\ref{betac-final}).
}
\label{results-etanu}
\begin{center}
\begin{tabular}{ccccccc}
$\beta_{\rm min}$ & $\chi^2_{\rm estr}$/DOF & $\chi^2_{\rm fit}$/DOF &
    $\eta\nu$  & $\chi^2_{\rm estr}$/DOF & $\chi^2_{\rm fit}$/DOF &
    $\eta\nu$ \\
\hline
 0.2750 & 37.1/22 & 16.2/15 & 0.0249(1+1) &
        &  &  \\
 0.2780 & 37.1/22 & 16.1/14 & 0.0249(2+1) &
       15.7/11 & 12.4/14 & 0.0253(2+1) \\
 0.2800 & 36.7/21 & 11.0/13 & 0.0246(2+1) &
       15.7/11 & 11.6/13 & 0.0252(2+1) \\
 0.2810 & 36.7/20 &  5.9/12 & 0.0243(2+1) &
       15.7/11 & 10.6/12 & 0.0251(3+1) \\
 0.2820 & 36.3/19 &  5.5/11 & 0.0242(3+1) &
       15.7/11 & 11.0/11 & 0.0250(3+1) \\
 0.2830 & 35.1/17 &  4.0/10 & 0.0241(4+1) &
       15.5/10 &  9.9/10 & 0.0248(4+2) \\
 0.2835 & 33.2/15 &  1.5/ 9 & 0.0236(5+2) &
       15.5/9 &  4.9/9 & 0.0243(6+2) \\
 0.2840 & 22.6/13 &  4.7/ 8 & 0.0236(7+2) &
       12.0/8 &  2.7/8 & 0.0245(8+1) \\ 
\end{tabular}
\end{center}
\end{table} 

\begin{table}
\caption{Results for the critical exponent $\eta\nu$ using 
scaling corrections with exponent $\omega = 0.8$.
We report two different $\chi^2$. 
The first one ($\chi^2_{\rm estr}$) refers to the fit that allows the 
determination of the curve $F_\zeta(z)$, cf. Eq.~(\protect\ref{S2L-L-corr}), 
the second one ($\chi^2_{\rm fit}$) to the fit 
(\protect\ref{fit-Z-omega}). Beside the $\chi^2$ we also report 
the number of degrees of freedom (DOF). We only report two errors:
the first one is the statistical error, 
the second one gives the variation of the estimate as $\beta_c$ is 
varied within one error bar, cf. Eq.~(\protect\ref{betac-final}).
}
\label{results-etanu-corr}
\begin{center}
\begin{tabular}{cccc}
$\beta_{\rm min}$ & $\chi^2_{\rm estr}$/DOF & $\chi^2_{\rm fit}$/DOF &
    $\eta\nu$ \\
\hline
 0.2750 & 24.5/29 & 174/15 & 0.0179(2+1) \\
 0.2780 & 24.5/28 & 55.8/14 & 0.0193(2+1) \\
 0.2800 & 24.1/26 & 29.0/13 & 0.0203(3+1) \\
 0.2810 & 22.5/24 & 19.3/12 & 0.0210(4+2) \\
 0.2820 & 20.0/22 & 11.0/11 & 0.0217(5+2) \\
 0.2830 & 19.4/19 &  6.1/10 & 0.0228(7+2) \\
 0.2835 & 19.2/16 &  4.8/9 & 0.0226(8+2) \\  
 0.2840 & 14.2/13 &  5.3/8 & 0.0230(12+3) \\ 
\end{tabular}
\end{center}
\end{table}

The second analysis we consider uses the data at the critical point. 
For $\beta\approx \beta_c$ we have 
\begin{equation}
\zeta(\beta,L) = L^{-\eta} f_1(t L^{1/\nu}) + 
             L^{-\eta-\omega} f_2(t L^{1/\nu}),
\end{equation}
where $t \equiv \beta_c - \beta$. If we expand for $t\approx 0$ we can write 
\begin{equation}
\ln \zeta(\beta,L) = - \eta \ln L + a + b t L^{1/\nu} + c L^{-\omega}.
\end{equation}
We first perform an analysis neglecting scaling corrections ($c = 0$) 
for several values of $L_{\rm min}$.
If we fix $\beta_c = 0.285744(2)$ and $\nu = 0.683(3)$ we obtain:
\begin{itemize}
\item[]
$L_{\rm min} = 16$: $\eta = 0.0331(2+6)$, $\chi^2$/DOF = 76/26; 
\item[]
$L_{\rm min} = 32$: $\eta = 0.0347(3+8)$, $\chi^2$/DOF = 27/21; 
\item[] 
$L_{\rm min} = 64$: $\eta = 0.0354(6+14)$, $\chi^2$/DOF = 22/13.
\end{itemize}
The first error is statistical, while the second one is due to the error on 
$\beta_c$. The error due to the error on $\nu$ can be neglected. 
There are apparently some scaling 
corrections, but it should be noted that the difference between the results 
with $L_{\rm min} = 32$ and 64 is not significant given the statistical error
bars. We also perform an analysis with scaling corrections with exponent 
$\omega = 0.80$. Using all data, i.e. taking $L_{\rm min} = 16$, we obtain 
$\eta = 0.0382(8+16)$, $c = -0.111(16+23)$, $\chi^2$/DOF = 29/25. 
This result is higher than the estimates obtained before, but still 
compatible with the quoted errors.

The analysis at the critical point gives therefore results that are in full 
agreement with those obtained before, confirming the correctness of 
the estimate (\ref{eta-final}), and with the result of Ref. \cite{BFMMPR-98},
$\eta = 0.0374(45)$.

\subsection{The exponent $\alpha$ and hyperscaling} \label{secIIF}

In this Section we wish to determine the exponent $\alpha$ and check
the hyperscaling relation $2-\alpha = d\nu$. Unfortunately, we have 
measured the specific heat $C(\beta,L)$ only near the critical point
and thus we can determine $\alpha$ only from the behavior for 
$\beta\approx \beta_c$. 

RG predicts that, in the FSS limit, the energy scales as
\begin{equation}
E(\beta,L) \approx E_{\rm bulk}(\beta) + L^{(\alpha-1)/\nu} g_1(t L^{1/\nu}) + 
        L^{(\alpha-1)/\nu-\omega} g_2(t L^{1/\nu}),
\label{Energy-FSS}
\end{equation}
where $t \equiv \beta_c - \beta$ and we have included one 
scaling correction. Note that $E_{\rm bulk}(\beta)$ is expected to have 
an exponentially small dependence on $L$ which can be neglected for all 
practical purposes. Near the critical point we can expand this expression
in powers of $t L^{1/\nu}$ 
obtaining for the energy and the specific heat the expressions
\begin{eqnarray}
E(\beta,L) &\approx& a_E + b_E t + c_E L^{(\alpha-1)/\nu} + 
    d_E t L^{\alpha/\nu} + e_E L^{(\alpha-1)/\nu-\omega},
\label{E-FSS}\\
C(\beta,L) &\approx& 
   a_C + b_C t + c_C L^{\alpha/\nu} + d_C t L^{(\alpha + 1)/\nu} + 
     e_C L^{\alpha/\nu-\omega},
\label{C-FSS}
\end{eqnarray}                                                
with $a_C = b_E$ and $c_C = d_E$. 

We wish now to determine $\alpha/\nu$. We consider the specific heat, 
since in this case the singular behavior is stronger. In Eq.~(\ref{C-FSS})
we have not used hyperscaling, so that the expression depends on 
two independent exponents $\alpha$ and $\nu$, 
or more precisely $\alpha/\nu$ and $1/\nu$. 
In order to simplify the analysis, we now analyze the specific heat 
using hyperscaling to rewrite the correction term. Thus, 
Eq.~(\ref{C-FSS}) becomes
\begin{eqnarray}
C(\beta,L) &=& a_C + b_C t + c_C L^{\alpha/\nu} + d_C t L^{3/2(\alpha/\nu + 1)}
  + e_C L^{\alpha/\nu-\omega},
\end{eqnarray}
that depends only on $\alpha/\nu$. We begin by neglecting the 
correction-to-scaling term, i.e. we set $e_C = 0$. 
Fixing $\beta_c = 0.285744(2)$, a five-parameter fit of the data with 
$L\ge L_{\rm min}$ gives:
\begin{itemize}
\item[] $L_{\rm min} = 16$, $\alpha/\nu = -0.119(7+6)$, $\chi^2$/DOF = 12.0/24;
\item[] $L_{\rm min} = 32$, $\alpha/\nu = -0.115(17+11)$, 
       $\chi^2$/DOF = 11.6/19.
\end{itemize}
The first error is the statistical one, while the second one gives the 
variation of the estimate as $\beta_c$ is varied by one error bar. 
Two things should be noticed: the analytic term proportional to $t$ is 
very small compared with the statistical errors. Indeed, 
$b_C = 23(83+8)$ and $b_C = -30(167+24)$ in the two fits. Moreover, corrections
to scaling are apparently small, since the results do not depend 
on $L_{\rm min}$. As a check, we have also performed an analysis 
with a correction term ($e_C\not= 0$). We find $\alpha/\nu = -0.109(50+6)$,
$e_C = -0.3(1.4)$, and $\chi^2$/DOF = 11.9/23, for $\omega = 0.8$ and 
$L_{\rm min} = 16$. The coefficient $e_C$ is compatible with zero, while the 
$\chi^2$ of the fit changes only by 0.1 in spite of the fact that we 
have added an additional parameter. There is no evidence of 
scaling corrections. As final estimate of $\alpha/\nu$ we take
the result with $L_{\rm min} = 32$ and $e_C = 0$, 
\begin{equation}
{\alpha\over \nu} = - 0.115(28).
\label{alphasunu-stima}
\end{equation}
If we use hyperscaling and the estimate of $\nu$ of Sec.~\ref{secIID}, 
$\nu = 0.683(3)$, we obtain $\alpha/\nu = - 0.072(13)$. 
This estimate is in reasonable agreement with (\ref{alphasunu-stima}) 
confirming the validity of hyperscaling. 
Using Eq.~(\ref{alphasunu-stima}) and $\nu = 0.683(3)$ we obtain 
$\alpha = -0.079(19)$,
which is in reasonable agreement with the estimate obtained using hyperscaling
$\alpha = -0.049(9)$.

We can also check hyperscaling directly by comparing the results for the 
energy and the specific heat. If we define 
$\theta_1 \equiv \alpha/\nu$, $\theta_2 \equiv (2-\alpha)/\nu$, 
we can rewrite Eqs.~(\ref{E-FSS}) and (\ref{C-FSS}) as 
\begin{eqnarray}
E(\beta,L) &=& a_E + b_E t + c_E L^{(\theta_1 - \theta_2)/2}  +
               d_E t L^{\theta_1}, \nonumber \\
C(\beta,L) &=& b_E + d_E L^{\theta_1} + d_C t L^{(3 \theta_1 + \theta_2)/2},
\label{fit-E-C}
\end{eqnarray} 
where we have neglected scaling corrections and we have set $b_C = 0$ 
in agreement with the previous analysis. If hyperscaling is satisfied 
we should find $\theta_2 = 3$. A combined analysis of $E(\beta,L)$ and 
$C(\beta,L)$ fixing $\beta_c$ gives
\begin{itemize}
\item[]
$L_{\rm min} = 16$: $\alpha/\nu = -0.119(6+6)$, $\theta_2 = 2.952(17+8)$,
         $\chi^2$/DOF = 33.4/51; 
\item[]
$L_{\rm min} = 32$: $\alpha/\nu = -0.112(16+9)$, $\theta_2 = 2.930(43+12)$,
  $\chi^2$/DOF = 29.3/41.
\end{itemize}
Here, as before, the first error is the statistical one, while the second 
gives the error due to $\beta_c$. The estimates of $\alpha/\nu$ are 
compatible with Eq.~(\ref{alphasunu-stima}), while $\theta_2$ is 
in reasonable agreement with 3, confirming again the validity of 
hyperscaling.

\subsection{The universal ratio $R_\xi^+$} \label{secIIG}

\begin{figure}[t]
\centerline{\psfig{width=8truecm,angle=-90,file=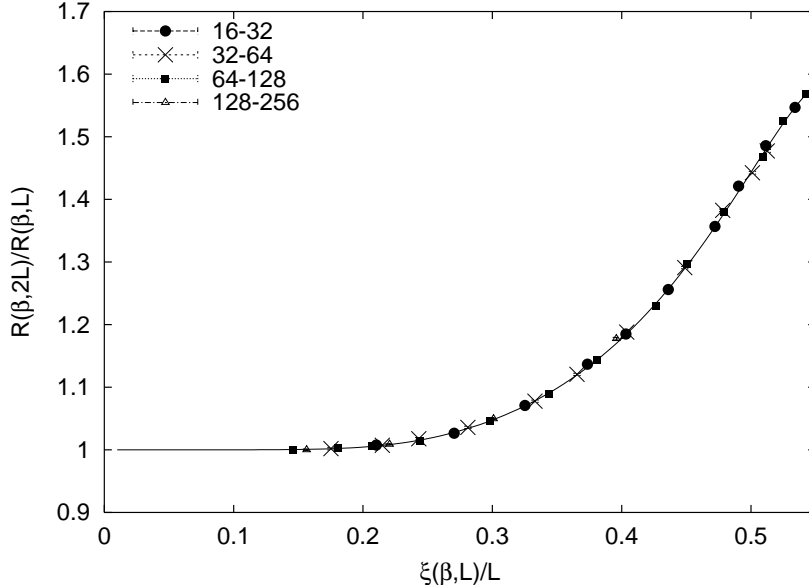}}
\caption{Ratios $R(\beta,2L)/R(\beta,L)$ vs $\xi(\beta,L)/L$.
The solid curve is a fit using all data with $L\ge 64$.
}
\label{figFSSRxi}
\end{figure} 

We now wish to compute the universal ratio $R_\xi^+$ that is related to 
the universality of the singular part of the free energy per correlation volume.
More precisely, let us consider the infinite-volume free energy density 
${\cal F}_\infty(\beta)$
\begin{equation}
{\cal F}_\infty = - {1\over V} \overline{\ln Z},
\end{equation}
and the infinite-volume specific heat 
$C_\infty(\beta) = {\partial^2 {\cal F}_\infty(\beta)/ \partial\beta^2}$. 
In the critical limit $t\equiv \beta_c - \beta \to 0$ we can write
\begin{eqnarray}
{\cal F}_\infty(\beta) = 
 {\cal F}_{\rm bulk}(\beta) + F^+ t^{2-\alpha} + \hbox{\rm corrections},
\nonumber \\
C_\infty(\beta) = 
C_{\rm bulk}(\beta) + A^+ t^{-\alpha} + \hbox{\rm corrections},
\end{eqnarray}
where $A^+ = (2 -\alpha)(1-\alpha)F^+$. If 
$\xi_\infty(\beta) \approx f^+ t^{-\nu}$ in the same limit $t\to 0^+$, 
RG predicts the 
universality of $F^+ (f^+)^3$, or equivalently of 
\begin{eqnarray}
R^+_\xi \equiv \left(\alpha A^+\right)^{1/3} f^+.
\end{eqnarray}
We now compute $R^+_\xi$ using our high-temperature results for the 
energy $E(\beta,L)$ and for the correlation length $\xi(\beta,L)$. 
We define a quantity $R(\beta,L)$,
\begin{equation}
R(\beta,L) \equiv \left[-{\alpha\over 1 - \alpha} 
       (E(\beta,L) - a_E - b_E t) t\right]^{1/3} \xi(\beta,L),
\label{definition-R}
\end{equation}
where  $a_E$ and $b_E$ are defined in terms of the expansion 
of ${\cal F}_{\rm bulk}(\beta)$ for $t\to 0$:
\begin{equation} 
   {d{\cal F}_{\rm bulk}(\beta)\over d\beta} \equiv 
   E_{\rm bulk}(\beta) =  a_E + b_E t + O(t^2).
\label{Ebulk}
\end{equation}
It is easy to check that 
\begin{equation}
\lim_{t\to 0} \lim_{L\to\infty} R(\beta,L) = R^+_\xi.
\end{equation}
In order to compute $R(\beta,L)$, we must specify the values of the two 
constants $a_E$ and $b_E$ in Eq.~(\ref{definition-R}). 
For this purpose we exploit the fact that 
$E_{\rm bulk}(\beta)$ is the same function in Eq.~(\ref{Energy-FSS}) and 
in Eq.~(\ref{Ebulk}), so that 
$a_E$ and $b_E$ coincide with the constants defined in Eq.~(\ref{E-FSS}).
Thus, $a_E$ and $b_E$ can be determined independently by 
using the critical-point data 
for the energy and the specific heat. We thus proceed as follows.
We consider Eq.~(\ref{fit-E-C}), fix $\beta_c = 0.285744(2)$, 
$\theta_2 = 3$, $\alpha/\nu = 2/\nu - 3$, $\nu = 0.683(3)$, and compute 
$a_E$ and $b_E$ by analyzing $E(\beta,L)$ and $C(\beta,L)$ near the 
critical point. Then, we determine $R(\beta,L)$. The error on $R(\beta,L)$ 
takes into account the error on
$E(\beta,L)$, $\xi(\beta,L)$, $a_E$, and $b_E$, and also the 
variation of the estimates as $\nu$ and $\beta_c$ vary within one error bar. 
In order to be conservative, we use a worst-error estimate summing all
errors together. Once $R(\beta,L)$ has been computed we use the extrapolation
method presented in Sec.~\ref{secIIC}. 

In Fig.~\ref{figFSSRxi} we report the ratios $R(\beta,2L)/R(\beta,L)$ together
with a fit of the data with $L\ge 64$ (a good fit is obtained by 
using a polynomial with $n=10$).
Apparently, there are no scaling corrections, but at a closer 
look one finds systematic deviations for $L = 16$. As before, these points 
will be discarded in the analytic fits. 

\begin{table}
\caption{Results for the universal ratio $R^+_\xi$.
Definitions are as in Table \ref{results-G4}. The error due to $\beta_c$ is 
negligible.  }
\label{results-Rxi}
\begin{center}
\begin{tabular}{ccccccc}
$\beta_{\rm min}$ & $\chi^2_{\rm estr}$/DOF & $\chi^2_{\rm fit}$/DOF &
    $R^+_\xi$  & $\chi^2_{\rm estr}$/DOF & $\chi^2_{\rm fit}$/DOF &
    $R^+_\xi$ \\
\hline
0.2750 &  14.7/19 & 40.2/16 & 0.28674(13) &
       &  & \\
0.2780 &  14.7/19 & 31.3/15 & 0.28688(14) &
      10.4/8 & 20.1/15 & 0.28677(16)\\
0.2800 &  14.6/18 & 23.9/14 & 0.28703(15) &
      10.4/8 & 16.5/14 & 0.28689(17)\\
0.2810 &  14.6/17 & 20.9/13 & 0.28712(17) &
      10.4/8 & 15.8/13 & 0.28698(19)\\
0.2820 &  13.8/16 & 20.4/12 & 0.28718(18) &
      10.4/8 & 13.2/12 & 0.28709(20)\\
0.2830 &  13.6/14 & 15.4/11 & 0.28744(22) &
      10.3/7 & 10.8/11 & 0.28726(23)\\
0.2835 &  12.9/12 & 12.8/10 & 0.28749(26) &
      10.3/6 & 9.45/10 & 0.28736(27)\\
0.2840 &  12.9/10 & 11.5/9 & 0.28737(32) &
      10.2/5 & 8.63/9 & 0.28730(33)\\
\end{tabular}
\end{center}
\end{table}

\begin{table}
\caption{Results for the universal ratio $R^+_\xi$ using 
scaling corrections with exponent $\omega = 0.8$.
Definitions are as in Table \ref{results-G4-G22-corr}. 
The error related to $\beta_c$ is negligible. }
\label{results-Rxi-corr}
\begin{center}
\begin{tabular}{cccc}
$\beta_{\rm min}$ & $\chi^2_{\rm estr}$/DOF & $\chi^2_{\rm fit}$/DOF &
    $R^+_\xi$ \\
\hline
0.2750 & 32.0/25 & 10.5/16& 0.28908(19)\\
0.2780 & 30.8/24 & 9.5/15 & 0.28897(21)\\
0.2800 & 30.5/22 & 8.6/14 & 0.28895(24)\\
0.2810 & 29.7/20 & 7.6/13 & 0.28897(26)\\
0.2820 & 28.5/18 & 8.5/12 & 0.28898(30)\\
0.2830 & 27.9/15 & 8.5/11 & 0.28905(36)\\
0.2835 & 17.8/12 & 5.2/10 & 0.28951(45)\\
0.2840 & 15.7/9  & 6.4/9  & 0.28967(56)\\
\end{tabular}
\end{center}
\end{table}

The results of the fits with analytic corrections are reported in Table
\ref{results-Rxi} for $L_{\rm min} = 32$ and $L_{\rm min} = 64$. For small 
$\beta_{\rm min}$ they show an upward trend and then apparently stabilize 
around $0.2874(3)$. In order to check the role of the corrections to
scaling, we have  
repeated the analysis by adding scaling corrections with exponent 
$\omega = 0.80$. The results are presented in Table \ref{results-Rxi-corr}.
They are now substantially independent of $\beta_{\rm min}$ confirming
that the data are very well fitted by assuming such an exponent.
The final estimate is somewhat higher than that obtained in the analytic fits,
indicating that in this case nonanalytic scaling correction may play an 
important role. We do not know which of the two fits is the most reliable 
one and thus we have taken as final estimate
\begin{equation}
R^+_\xi = 0.2885(15),
\label{Rxi-final}
\end{equation}
which is compatible with the results of both analyses.

The estimate (\ref{Rxi-final}) is in good agreement with the 
results of other methods. A six-loop computation in the fixed-dimension
FT approach gives \cite{CPV-03-eqstate} $R^+_\xi = 0.290(10)$, 
while, by using approximate parametric representations of the 
equation of state, one obtains \cite{CPV-03-eqstate} $R^+_\xi = 0.282(3)$.

\section{Comparison with field-theory results}
\label{fieldtheory}

The critical behavior of the RIM has been extensively studied 
using the FT approach. 
Quantitative predictions can be obtained by using different techniques:
perturbative methods in the four-point renormalized couplings in 
fixed dimension $d=3$ or in  $\sqrt{\epsilon}$, $\epsilon\equiv 4 - d$, or 
nonperturbative methods based on approximate RG
equations, see 
Refs.~\cite{SAS-97,FHY-00,BH-03,PV-00,CPPV-03,PSS-02,%
HY-98,FHY-99,Varnashev-00,PS-00,TMVD-02}
The most accurate results have been obtained in the first approach: 
six-loop expansions
for the $\beta$-functions and the critical exponents have been 
derived and analyzed in Refs.~\cite{CPV-00,PV-00}. The corresponding 
estimates of the critical exponents, e.g., $\nu=0.678(10)$ and $\eta=0.030(3)$,
are in satisfactory agreement with the Monte Carlo results presented before.

The main problem of the perturbative approach is the non-Borel
summability of the series \cite{BMMRY-87,AMR-99}.
This fact makes  the analysis more subtle and less precise than
in the case of the pure Ising model. 
The difficulties of the perturbative approach 
appear in the determinations of the fixed-point values
$u^*$ and $v^*$ of the renormalized couplings (they are normalized so that 
at tree level $u = u_0/m$, $v = v_0/m$, $m$ being the renormalized mass), 
which are directly related to the quantities we have measured in the 
Monte Carlo simulation (see Ref.~\cite{CPV-03-eqstate} for a derivation of these
relations): $G^*_4 = v^*$, $G_{22}^* = u^*/3$, and 
$H_4^* \equiv G_4^* + 3 G_{22}^* = u^* + v^*$.
The analysis of the six-loop series provided results somewhat dependent on 
the resummation method \cite{foot-normalizzazioni}. 
Indeed, we found \cite{PV-00}: 
$G_4^* = 37.7(2)$, $G_{22}^* = -4.3(6)$, $H_4^* = 24.8(1.8)$ 
(double Pad\'e-Borel method); 
$G_4^* = 36.8(3)$, $G_{22}^* = -4.0(1)$, $H_4^* = 24.8(6)$ 
(conformal Pad\'e-Borel method);
$G_4^* = 38.6(7)$, $G_{22}^* = -4.8(2)$, $H_4^* = 24.2(9)$ 
(direct conformal method). The estimates of $H^*_4$ are in good agreement
with the Monte Carlo result (\ref{Hstar4-final}), 
$H_4^* = 24.7(2)$. On the other hand, the estimates of $G_4^*$  and 
$G_{22}^*$---combining the results we would have guessed 
$G_4^*=38.0(1.5)$ and $G^*_{22} = -4.5(6)$ with errors that are,
at first sight, quite conservative---differ significantly from the 
Monte Carlo estimates (\ref{Gstar4-final}) and (\ref{Gstar22-final}). 

These discrepancies call for a reanalysis of the perturbative series
of the exponents, verifying if the use of the Monte Carlo results
for $G_4^*$ and $G_{22}^*$  leads to significantly different  estimates.
We have thus repeated the analysis,  using the different resummation methods 
outlined in Refs.~\cite{PV-00,PS-00}.  We find
\begin{equation}
\nu=0.686(4),\qquad \eta=0.026(3),\qquad \gamma=1.355(8),
\label{expan}
\end{equation}
where the errors include the results of the different resummation methods.
It is reassuring that these estimates are close to those 
found in Ref.~\cite{PV-00}, 
$\nu=0.678(10)$, $\eta=0.030(3)$, and $\gamma=1.330(17)$, 
and also reasonably close to the Monte Carlo estimates. 
The small variation of the estimates of the critical exponents 
is due to the particular structure of the perturbative series: 
if they are rewritten in terms of $y \equiv u + v$ and $u$,
the resummations depend mostly on $y^*= H_4^*$, which is correctly determined 
by FT methods, 
and only slightly on $u^*$ that is instead poorly known.
We should also observe that the new estimate of $\nu$ is closer to the Monte 
Carlo result, while the estimate of $\eta$ is slightly worse. 
Therefore, the FT estimates do not become more accurate if
more precise results for $u^*$ 
and $v^*$ are used. This is an indication that, at least for the critical exponents,
the location of the fixed point is not 
the main source of error on the results.

\begin{table}[tbp]
\caption{
Estimates of the coefficients 
$\bar{c}_{ij} = (16 \pi/3)^i (6 \pi)^j c_{ij}(g_I^*)$, 
cf. Eq.~(\ref{cijdef}),
for the expansions of $\eta$, $\nu$, $1/\nu$, $\gamma$, and $1/\gamma$
around the Ising fixed point.
}
\label{cij}
\begin{tabular}{llllll}
\multicolumn{1}{c}{}&
\multicolumn{1}{c}{$\eta$}&
\multicolumn{1}{c}{$\nu$}&
\multicolumn{1}{c}{$1/\nu$}&
\multicolumn{1}{c}{$\gamma$}&
\multicolumn{1}{c}{$1/\gamma$}\\
\tableline 
$\bar{c}_{10}$ & 0.050(6)     & 0.105(3)     & $-$0.273(5) & 0.181(3) & $-$0.1193(8) \\
$\bar{c}_{20}$ & 0.032(8)     & 0.016(4)     &  0.00(1) & 0.014(5) & 0.008(6)\\
$\bar{c}_{30}$ & 0.011(6)     & 0.004(5)     &  0.00(3)& 0.00(1) & 0.00(1) \\

$\bar{c}_{01}$ & 0            & 0.0500(4)    & $-$0.127(1) & 0.0987(6) &$-$0.0646(1) \\
$\bar{c}_{02}$ & $-$0.0062(2) & 0.0056(9)    & $-$0.003(3)& 0.015(2) & $-$0.005(2)\\
$\bar{c}_{03}$ & 0.0010(2)    & $-$0.003(1)  & 0.017(3)& $-$0.007(2) & 0.06(1)\\
$\bar{c}_{04}$ & 0.0001(5)    & 0.00(1)      & 0.00(1)& 0.00(2) & 0.001(1)\\

$\bar{c}_{11}$ & 0            & 0.017(1)     & $-$0.003(1)& 0.032(1) & $-$0.0024(2)\\
$\bar{c}_{12}$ & $-$0.0018(4) & $-$0.003(2)  & 0.018(6)& $-$0.006(3) & 0.009(3)\\
$\bar{c}_{13}$ & 0.0007(4)    & $-$0.004(3)  & 0.00(1)& $-$0.009(7) & 0.005(5)\\

$\bar{c}_{21}$ & 0            & 0.0033(6)    & 0.0006(6)& 0.005(1) & $-$0.0004(2)\\
$\bar{c}_{22}$ & $-$0.0009(4) & $-$0.003(2)  & 0.007(1)& $-$0.005(4) & 0.003(1)\\
\end{tabular}
\end{table}

We also tried an alternative procedure based on an expansion of the RG 
functions around the unstable Ising fixed point $u=0$, $v=g_I^*$, 
where \cite{CPRV-02} $g_I^*=23.56(2)$.
The analysis of the Ising-to-RIM RG flow 
reported in Ref.~\cite{CPPV-03} and the discussion reported above show that it
is convenient to introduce new variables $y\equiv u+v$ and $z\equiv -u$.
In terms of $y$ and $z$, the RIM fixed point is located in $y^*=H_4^*=24.7(2)$
and $z^*=-3 G_{22}^* = 18.6(3)$, while the Ising fixed point
is at $y_I=g_I^*$, $z=0$.
Then, we write $y = g_I + \delta y$, obtaining for any RG function 
$f(y,z)$,
\begin{eqnarray}
&& f({y},{z})=
\sum_{i,j}  c_{ij}(g_I) \delta{y}^i\, {z}^j,
\label{expansion1-f} \\
&& c_{ij}(g_I)=\sum_k f_{ijk}\, {g}_{I}^k.
\label{cijdef}
\end{eqnarray}
The value of $f(y,z)$ at the RIM fixed point is then obtained as follows.
First, we compute the coefficients $c_{ij}(g^*_I)$ at the Ising 
fixed point, by  using the conformal-mapping
method and exploiting the known large-order
behavior of the expansion of $c_{ij}(g_I)$ that is determined
by the Ising fixed point. 
Then, we evaluate the double series at
$\delta{y}^* = H^*_4 - g^*_I = 1.14(20)$ and $z^* = -3 G_{22}^* = 18.6(3)$.
Here, we are neglecting the fact that the RG functions are nonanalytic
at the Ising fixed point \cite{Nickel-81,PV-98,CCCPV-00,CPV-Kleinert}.
Note that $f({g}_{I}^*,0)$ is the value of the same quantity
for the Ising universality class, so that
the expansion (\ref{expansion1-f}) provides
the differences between RIM and Ising critical exponents,
i.e. $\Delta f = f_{\rm RIM} - f_{\rm Ising}$,
which are expected to be rather small.
Of course, this expansion is, at most, asymptotic.
But one may hope that the RIM and Ising fixed points are sufficiently
close, so that the first few terms of the expansion around the Ising fixed 
point  allow us to obtain more accurate
estimates of the exponents of the RIM.
In Table~\ref{cij} we report the 
estimates of the first coefficients $c_{ij}(g_I^*)$
for $\eta-\eta_I$, $\nu-\nu_I$, $\gamma-\gamma_I$, 
$1/\nu-1/\nu_I$, and $1/\gamma-1/\gamma_I$.
The results for the critical exponents are reported in Table~\ref{resexp}
as a function of the order $o\equiv{\rm Max}[i+j]$ of the expansion.
We thus obtain the following estimates:
$\eta-\eta_I=-0.0017(13)$,
$\nu-\nu_I=0.060(5)$,
$1/\nu-1/\nu_I= -0.135(15)$,
$\gamma-\gamma_I=0.12(1)$,
$1/\gamma-1/\gamma_I=-0.07(1)$.
The estimate we quote corresponds to $o=3$, while the error is such to include 
the results with $o=2$ and $o=4$.
Then, using the estimates \cite{PV-r}
$\eta_I=0.0364(5)$, $\nu_I=0.6301(4)$, $\gamma_I=1.2372(5)$,
we finally obtain
\begin{equation}
\nu=0.690(5),\qquad \eta=0.035(2),\qquad \gamma=1.357(10),
\label{expexp}
\end{equation}
which are again in substantial agreement with the MC results.

\begin{table}[tbp]
\caption{Results obtained by using the  expansion (\protect\ref{expansion1-f})
for various truncations $o\equiv{\rm Max}[i+j]$. The first 
error is due to the uncertainty on the values of the 
coefficients $c_{ij}(g_I^*)$, the second one is due to the 
uncertainty on the location of the RIM fixed point.}
\label{resexp}
\begin{tabular}{clllll}
\multicolumn{1}{c}{$o$}&
\multicolumn{1}{c}{$\eta-\eta_I$}&
\multicolumn{1}{c}{$\nu-\nu_I$}&
\multicolumn{1}{c}{$1/\nu-1/\nu_I$}&
\multicolumn{1}{c}{$\gamma-\gamma_I$}&
\multicolumn{1}{c}{$1/\gamma-1/\gamma_I$}\\
\tableline 
1 &0.0033(4+6) & 0.056(0+2) & $-$0.145(0+5) & 0.110(1+4) & $-$0.072(1+2)\\
2 &$-$0.0025(4+8) &0.063(1+2) & $-$0.147(3+5) & 0.126(2+5) &$-$0.076(1+3)\\
3 &$-$0.0017(5+8) &0.060(1+2) & $-$0.135(3+5) & 0.120(2+4) &$-$0.070(2+2)\\
4 &$-$0.0016(6+8) &0.062(10+3)& $-$0.137(20+5)& 0.123(20+4)&$-$0.070(10+2)\\
\end{tabular}
\end{table}

The expansion around the Ising fixed point can also be performed along the
Ising-to-RIM RG trajectory \cite{CPPV-03}, which is obtained as the 
limit $u_0\rightarrow 0^-$ of the RG trajectories in the $u$, $v$ plane.
At least in principle, this expansion is expected to be better behaved than
the previous one, since RG functions should be analytic near the 
Ising fixed point only along this trajectory, up to the random fixed point
where nonanalyticites are again present \cite{CPPV-03}.

An effective parametrization of the curve is given by the first few terms of 
its expansion
around $z=0$, which is given by
\begin{equation}
y - y_I = T(z) = c_2 z^2 + c_3 z^3 + ...
\label{expansion-T}
\end{equation}
where\cite{CPPV-03} $c_2=0.0033(1)$ and $c_3=1(2)\times 10^{-5}$.
The fact that $y-y_I$ is of order $z^2$ is the main reason why 
we introduced the variable $y$ and is due to the identity 
\cite{CPPV-03}
\begin{eqnarray}
&&
\left. {\partial \beta_v\over \partial u} \right|_{u=0}
+ \left. {\partial \beta_u\over \partial u} \right|_{u=0}
- \left. {\partial \beta_v\over \partial v} \right|_{u=0}
=0.
\end{eqnarray} 
Substituting the expansion (\ref{expansion-T}) into the double expansion
(\ref{expansion1-f}), we obtain an expansion in powers of $z$
\begin{equation}
f(y,z)  =f({g}_{I}+T({z}),{z})= \sum_i e_i(g_I) {z}^i,
\label{eidef}
\end{equation}
which should be then evaluated
at $g_I = g_I^*$ and $z=z^*$.
The values of the coefficients $e_i(g_I^*)$ at the Ising fixed point 
have been computed by using a conformal mapping and a Borel 
transform. The results for $i=1,2,3$ are reported in Table \ref{ei}.
The estimates of the difference between 
the critical exponents of the Ising and RIM universality classes
are 
$\eta-\eta_I=-0.0020(18)$,
$\nu-\nu_I=0.060(5)$,
$1/\nu-1/\nu_I= -0.136(20)$,
$\gamma-\gamma_I=0.119(10)$, and
$1/\gamma-1/\gamma_I=-0.070(10)$.
These results are obtained by truncating the expansion (\ref{eidef}) 
to third order, while 
the error is the sum of the uncertainty
due to the resummation, due to the truncation of the series
(the difference between  the second-order and the third-order result), 
and due to the uncertainty on $z^*$ (we used the Monte Carlo result). 
Then, by using the estimates \cite{PV-r}
$\eta_I=0.0364(5)$, $\nu_I=0.6301(4)$, $\gamma_I=1.2372(5)$,
we finally obtain
\begin{equation}
\nu=0.690(8),\qquad \eta=0.0345(20), \qquad \gamma=1.355(10),
\label{expexp2}
\end{equation}
that do not differ significantly from the estimates (\ref{expexp}). 

Note that the estimate of $\eta$ obtained by using the expansion 
around the Ising fixed point is now in perfect agreement with 
the Monte Carlo result, at variance with the direct estimate
(\ref{expan}). The estimate of $\nu$ is also in substantial 
agreement with the numerical estimate $\nu = 0.683(3)$. Therefore, 
the expansion around the Ising fixed point appears to be a 
useful alternative method to compute the critical properties of the RIM.

\begin{table}[tbp]
\caption{
Estimates of the coefficients 
$\bar{e}_{i} = (6\pi)^i e_{i}(g_I^*)$, cf. Eq.~(\ref{eidef}),
for the expansions of $\eta$, $\nu$, $1/\nu$, $\gamma$, and $1/\gamma$
around the Ising fixed point.
}
\label{ei}
\begin{tabular}{llllll}
\multicolumn{1}{c}{$i$}&
\multicolumn{1}{c}{$\eta$}&
\multicolumn{1}{c}{$\nu$}&
\multicolumn{1}{c}{$1/\nu$}&
\multicolumn{1}{c}{$\gamma$}&
\multicolumn{1}{c}{$1/\gamma$}\\
\tableline 
1 & 0           & 0.0500(6)&   $-$0.1278(4) & 0.0987(6) &   $-$0.06462(7)\\
2 &$-$0.0028(4) &   0.013(1) &$-$0.022(3)  &   0.027(2)  &$-$0.013(2)  \\
3 & 0.0008(4) &   $-$0.002(1) & 0.012(2)  &   $-$0.005(2)  & 0.0065(8) \\
\end{tabular}
\end{table}

\section*{Acknowledgments}

V.M.-M. is a {\em Ramon y Cajal} research fellow and is partly supported by 
MCyT (Spain), project Nos. FPA2001-1813 and FPA2000-0956.
The numerical computations were performed at the PC Cluster at the 
University of Pisa. We thank Maurizio Davini for his
indispensable technical assistance.

\appendix

\section{Notations}
\label{AppA}

We consider the Hamiltonian (\ref{latticeH}) with $J = 1$ 
on a finite lattice $L^3$ with periodic boundary conditions.
Given a quantity ${\cal O}$ depending on the spins $\{s\}$ and on the 
random variable $\{\rho\}$ we define the sample average at fixed 
distribution $\{\rho\}$ 
\begin{equation}
\langle {\cal O} \rangle (\beta,\{\rho\}) \equiv 
    {1\over Z(\{\rho\})} \sum_{\{s_i\}} {\cal O} e^{-\beta {\cal H}[s,\rho]},
\end{equation}
where $Z(\{\rho\})$ is the sample partition function. Of course, we are 
interested in averaging over the random dilution and thus we consider 
disorder-averaged quantities
\begin{equation}
\overline{\langle {\cal O} \rangle} (\beta) = 
   \int [d\rho] \langle {\cal O} \rangle (\beta,\{\rho\}) ,
\end{equation}
where 
\begin{equation}
[d\rho] = \prod_i [x \delta(\rho_i - 1) + (1-x) \delta(\rho_i) ].
\end{equation}
We define the two-point correlation function and the susceptibility 
$\chi(\beta,L)$ 
\begin{eqnarray}
G(x;\beta,L) &\equiv& \overline{\langle \rho_0 s_0 \,\rho_x s_x \rangle}
\\
\chi(\beta,L) &\equiv& \sum_x G(x;\beta,L) 
\end{eqnarray}
We also consider the second-moment correlation length $\xi$.
In infinite volume it is defined as
\begin{equation}
\xi^2_\infty(\beta) \equiv  {1\over 6 \chi_\infty(\beta)} 
      \sum_x |x|^2 G_\infty(x;\beta).
\end{equation}
The finite-volume generalization is by no means unique.
We use
\begin{equation}
\xi^2(\beta,L) \equiv {\hat{G}(0;\beta,L) - \hat{G}(q_{\rm min};\beta,L) \over 
          \hat{q}_{\rm min}^2 \hat{G}(q_{\rm min};\beta,L) },
\end{equation}
where $q_{\rm min} \equiv (2\pi/L,0,0)$, $\hat{q} \equiv 2 \sin q/2$, and 
$\hat{G}(q;\beta,L)$ is the Fourier transform of $G(x;\beta,L)$. 
This finite-volume definition has the correct infinite-volume limit
and shows a fast convergence as $L\to\infty$
\cite{CP-98,foot-corrlength}.

We also define the energy $E(\beta,L)$ and the specific heat $C(\beta,L)$:
\begin{eqnarray}
E(\beta,L) &\equiv& 3 G(e;\beta,L), \nonumber \\
C(\beta,L) &\equiv& {\partial E(\beta,L)\over \partial \beta},
\end{eqnarray}
where $e = (1,0,0)$.
We also compute higher-order couplings. Setting
\begin{equation}
\mu_{k} \equiv \langle \; ( \sum_i \rho_i s_i\; )^k \rangle, 
\qquad
m_{k_1k_2...k_n} \equiv \overline{ \mu_{k_1} \mu_{k_2} ...\mu_{k_n}},
\end{equation}
we define the connected $n$-point susceptibilities $\chi_n$
averaged over random dilution by
\begin{eqnarray}
&&V \chi_2(\beta,L)\equiv V \chi(\beta,L) = m_2 , \\
&&V \chi_4(\beta,L) \equiv  m_4 - 3 m_{22}, 
\nonumber \\
&&V \chi_{6}(\beta,L) \equiv m_6 - 15 m_{42} + 30 m_{222},
\nonumber 
\end{eqnarray}
where $V\equiv L^3$ is the volume.
Moreover, we also define 
\begin{eqnarray}
&&V \chi_{22}(\beta,L) \equiv m_{22} - m_2^2, \nonumber \\
&&V \chi_{42}(\beta,L) \equiv m_{42} - m_4 m_2 - 3 m_{222} +  3 m_{22} m_2, 
\nonumber \\
&&V \chi_{222}(\beta,L) \equiv m_{222} - 3 m_{22} m_2 + 2 m_2^3.
\end{eqnarray}
Then, we define the four-point couplings 
\begin{eqnarray}
&&G_4(\beta,L) \equiv - {\chi_4 \over \xi^3 \chi_2^2}  ,
\nonumber \\
&&G_{22}(\beta,L) \equiv - {\chi_{22} \over \xi^3 \chi_2^2}  ,
\nonumber \\
&&H_4(\beta,L) \equiv G_4 + 3 G_{22} ,
\label{fourpoint-def}
\end{eqnarray}
and the six-point universal ratios
\begin{eqnarray}
&&r_6(\beta,L) \equiv 10 - {\chi_6 \chi_2 \over \chi_4^2},  \nonumber 
\\
&&C_{42}(\beta,L) \equiv 4 - {\chi_{42} \chi_2 \over \chi_4 \chi_{22}},
\nonumber \\
&&C_{222}(\beta,L) \equiv 6 - {\chi_{222}\chi_2 \over \chi_{22}^2}. 
\label{sixpoint-def}
\end{eqnarray}
We will be interested in the critical value of these quantities. 
If $S(\beta,L)$ is any of them, we compute its fixed-point value
(note that the order of the limits cannot be interchanged)
\begin{equation}
S^* = \lim_{\beta\to\beta_c} \lim_{L\to\infty} S(\beta,L).
\label{limSstar}
\end{equation}
Finally, we define the Binder parameters
\begin{eqnarray}
&& U_{n}(\beta,L) \equiv {m_{n}\over m_2^{n/2}}, \nonumber \\
&& U_{22}(\beta,L) \equiv  {m_{22}\over m_2^2},
\end{eqnarray}
and the corresponding critical-point values
\begin{equation}
U^* = \lim_{L\to\infty} \lim_{\beta\to\beta_c} U(\beta,L).
\label{limUstar}
\end{equation}
Note that the order of limits is reversed with respect to Eq.~(\ref{limSstar}).



\begin{references}
\bibitem{foot-omega-FT}
Perturbative field theory predicts $\omega = 0.25(10)$ (Ref.~\cite{PV-00}),
$\omega = 0.372(5)$ (Ref.~\cite{FHY-99}).
Moreover, it is possible to compute 
the next-to-leading correction exponent $\omega_2$. 
The analysis of the six-loop perturbative series of Ref.~\cite{PV-00} gives
$\omega_2 = 0.8(2)$ (Ref.~\cite{CPPV-03}).
\bibitem{foot-random}
The Parisi-Rapuano generator \cite{PR-85} is defined in terms of the sequence 
$a_{{\rm PR},n}$ defined by
$a_{{\rm PR},n} = {\rm mod}\, (a_{{\rm PR},n-24} + 
     a_{{\rm PR},n-55},2^{32})$. A random integer between 0 and $2^{32} - 1$ 
is obtained by taking $r_{{\rm PR},n} = a_{{\rm PR},n}\, {\tt XOR}\, 
a_{{\rm PR},n-61}$. We also consider the congruential generator
$r_{c,n} = {\rm mod}\, (16807 r_{c,n-1},2^{31}-1)$. 
The random numbers we use are obtained from
$r_n = {\rm mod}(2r_{c,n} + r_{{\rm PR},n},2^{32})$.
\bibitem{footJacobi}
From a practical point of view,
not any basis in the space spanned by the functions $\{ z^k \}$ is suitable to
perform the fit. As discussed in Ref. \cite{MPS-97},
care must be taken to choose functions which are
effectively orthogonal with respect to the scalar product induced by the
$\chi^2$, otherwise the numerical solution
of the minimization problem can be unstable.
We used the functions
$
f_k(z) = y P^{(0,5)}_{k-1}(2 y/y_{\text{max}}-1) ,
$
where $y \equiv e^{-1/z}$, $y_{\text{max}} \equiv e^{-1/z_{\rm max}}$, 
$P^{(a,b)}_n$ are Jacobi polynomials which are an orthonormal system in
$L^2([0,1],x^a (1-x)^b)$, and $z_{\text{max}}$ is the maximum value 
of $\xi(\beta,L)/L$ for the data 
(in our case $z_{\text{max}} \approx 0.542719$).
The choice of the parameters $a,b$ is
not crucial and different choices would perform analogously.
\bibitem{foot-normalizzazioni}
In Ref.~\cite{PV-00}, results are given for the rescaled couplings
$\bar{u}$ and $\bar{v}$. The relation is: $u \equiv 6 \pi \bar{u}$, 
$v \equiv 16 \pi \bar{v}/3$.
\end{references}
\end{document}